\documentclass[onecolumn,floats,showpacs,prd,amssymb,floatfix,nofootinbib,balancelastpage,superscriptaddress,amsmath,showkeys]{revtex4}
\usepackage{epsfig}

\def\simlt{\ \raise -2.truept\hbox{\rlap{\hbox{$\sim$}}\raise5.truept   %
\hbox{$<$}\ }}
\def\srunits{{\ {\rm cm}^{-2}{\rm s}^{-1}{\rm sr}^{-1}}}                %
\def\simgt{\ \raise -2.truept\hbox{\rlap{\hbox{$\sim$}}\raise5.truept   %
\hbox{$>$}\ }}                                                          %

\def\be{\begin{equation}}
\def\ee{\end{equation}}
\def\newline{\hfil\break}

\def\ergcm2s{{erg~cm$^{-2}$s$^{-1}$~}}

\def\la{\mathrel{\hbox{\rlap{\hbox{\lower4pt\hbox{$\sim$}}}\hbox{$<$}}}}
\def\ga{\mathrel{\hbox{\rlap{\hbox{\lower4pt\hbox{$\sim$}}}\hbox{$>$}}}}

\def\mug{$\mu$G ~}
\def\be{\begin{equation}}
\def\ee{\end{equation}}
\def\cm3{cm$^{-3}$}

\newcommand{\msun}{M_{\odot}{\ }}
\newcommand{\sun}{\odot}

\def\farcm{\hbox{$.\mkern-4mu^\prime$}}
\begin{document}

\title{Detecting dark matter WIMPs in the Draco dwarf: a multi-wavelength perspective}

\author{Sergio Colafrancesco}
\email{colafrancesco@mporzio.astro.it} \affiliation{INAF - Osservatorio Astronomico di
Roma,
              via Frascati 33, I-00040 Monteporzio, Italy and\\
              Istituto Nazionale di Fisica Nucleare,
              Sezione di Roma 2, I-00133 Roma, Italy}
\author{Stefano Profumo}
\email{profumo@caltech.edu}
\affiliation{Division of Physics, Mathematics and Astronomy,
              California Institute of Technology, \\
              Mail Code 106-38, Pasadena, CA 91125, USA}
\author{Piero Ullio}
\email{ullio@sissa.it}
\affiliation{Scuola Internazionale Superiore di Studi Avanzati,
              Via Beirut 2-4, I-34014 Trieste, Italy and\\
              Istituto Nazionale di Fisica Nucleare,
              Sezione di Trieste, I-34014 Trieste, Italy}


\begin{abstract}

We explore the possible signatures of dark matter pair annihilations in the nearby dwarf
spheroidal galaxy Draco. After investigating the mass models for Draco in the light of
available observational data, we carefully model the dark matter density profile, taking
advantage of numerical simulations of hierarchical structure formation. We then analyze
the gamma-ray and electron/positron yield expected for weakly interacting dark matter
particle (WIMP) models, including an accurate treatment of the propagation of the charged
particle species. We show that unlike in larger dark matter structures -- such as galaxy
clusters -- spatial diffusion plays here an important role. While Draco would appear as a
point-like gamma-ray source, synchrotron emission from electrons and positrons produced
by WIMP annihilations features a spatially extended structure. Depending upon the cosmic
ray propagation setup and the size of the magnetic fields, the search for a diffuse radio
emission from Draco can be a more sensitive indirect dark matter search probe than gamma
rays. Finally, we show that available data are consistent with the presence of a black
hole at the center of Draco: if this is indeed the case, very significant enhancements of
the rates for gamma rays and other emissions related to dark matter annihilations are
expected.

\end{abstract}




\maketitle

\section{Introduction}
 \label{Intro}

The astrophysical search for signals of dark matter (DM) particle pair annihilations in cosmic structures on large
scales (from galaxies to clusters of galaxies) is, potentially, a very powerful technique, highly complementary to direct DM searches, in the quest for the identification of the fundamental nature of DM. The widest and more definite set of results can be harvested through a multi-frequency
survey of DM annihilation signals over the whole electromagnetic (e.m.) spectrum (see, e.g.
\cite{Colafrancesco:2005ji}, hereafter CPU2006, and references therein) by using a detailed treatment of both the
microscopic interaction properties of the hadronic and leptonic secondary yields of WIMP annihilations, and of the subsequent emissions originating by the yields themselves in the astrophysical environment at hand. Various astrophysical systems have been taken into consideration to this aim. The central regions of ordinary galaxies (like our own Galaxy) are usually considered
among the best places to set constraints on the presence and on the nature of DM particles (see,
e.g, \cite{bertoneetal2004} for a review, and the analyses in \cite{cesarinietal2004,bergstrometal2005,baltzetal2002}, among others).

The typical faintness of DM signals within viable WIMP scenarios makes, in fact, the
Galactic Center, or the central regions of nearby galaxies (like M31), the most plausible and promsing places to detect signals of WIMP annihilations. However, the expected DM signals have to
contend, there, with the rich and often poorly understood astrophysical context of thermal and non-thermal sources (SN
remnants, pulsars, molecular clouds, to mention a few), whose spectral energy distributions
(SEDs) cover the whole e.m. spectrum, reaching even TeV energy scales, as the recent
results from HESS, MAGIC, Cangaroo have clearly shown (see, e.g., \cite{aharonian2006}
and references therein; see, however, also \cite{Profumo:2005xd}). In this respect, galaxy cores are likely not the best places to
definitely identify DM annihilation signals.

Galaxy clusters have the advantage to be mass-dominated by DM and, in some cases, like the
nearby Coma cluster, to have a quite extended spectral and spatial coverage of thermal
and non-thermal emission features which enable to set interesting constraints on the
properties of DM (see, e.g.,
\cite{ColafrancescoMele2001,Colafrancesco2004,Colafrancesco:2005ji} and refs. therein for
various aspects of the DM SEDs in clusters). The study of the DM-induced SEDs in galaxy
clusters has been shown to be quite constraining for DM WIMP models, and can even be advocated to
shed light on some emission features (e.g., radio halos, hard-X-ray and UV excesses, gamma-ray
emission) which are still unclear. Nonetheless, the sensitivity and spatial resolution of
the present and planned experiments in the gamma-rays, X-rays and radio do not likely
allow to probe more than a few nearby clusters. It is therefore, mandatory to remain within the local environment to have reasonable
expectations to detect sizable emission features of possible DM signals.

Globular clusters have also been proposed (see e.g., \cite{giraudetal2003}) as possible
sources of gamma rays from WIMP annihilations, but with expected signals well below the sensitivity threshold of
future experiments, mainly due to their quite low mass-to-light ratios.

The ideal astrophysical systems to be used as probes of the nature of DM
should be mostly dark (i.e., dominated by DM), as close as possible (in order to produce
reasonably high fluxes), and featuring central regions mostly devoid of sources of
diffuse radiation at radio, X-rays and gamma-rays frequencies, where the DM SEDs peak (see, e.g.
CPU2006 for general examples).

Dwarf spheroidal (dSph) galaxies closely respond to most of these requirements, as they generally
consist of a stellar population, with no hot or warm gas, no cosmic-ray population and
little or no dust (see, e.g., \cite{Mateo1998} for a review). Several dSph galaxies populate the region around the Milky Way and M31, and some of them seem to be
dynamically stable and featuring high concentrations of DM.

Among these systems, the Draco dSph is one of the most interesting cases. This object has already
been considered as a possible gamma-ray source fed by DM
annihilations in recent studies~\cite{Tyler2002,Evans:2003sc,BergstromHooper2006,ProfumoKam2006,Bietal2006}, in part triggered by an anomalous excess of photon counts from Draco reported by the CACTUS collaboration in a drift-scan mode survey of the region surrounding the dSph galaxy \cite{cactus}. The nature of the effect is still controversial, but it has been shown in \cite{ProfumoKam2006,BergstromHooper2006} to be in conflict, in most WIMP models, with the EGRET null-result in the search for a gamma-ray source from the direction of Draco \cite{way}. Other gamma-ray upper limits have been obtained by the Whipple 10-m telescope collaboration as well (see e.g., \cite{Vassilievetal2003}).

The observational state-of-the-art for Draco goes, however, beyond gamma-ray emissions: radio continuum upper limits on Draco have been obtained by Fomalont et al. \cite{Fomalontetal1979} with the VLA. These authors report an upper limit of $J_{\nu} < 2$ mJy at $\nu = 4.9$ GHz (this is a $3 \sigma$ level limit). Typical magnetic field strength of $B \sim 2-4$ \mug for dwarf galaxies similar to Draco have also been derived from radio observations at $5 $ GHz \cite{Klein1992}. The X-ray emission from the central part of Draco has an upper limit provided by ROSAT \cite{ZangMeurs2001}. The count rate detected by the PSPC instrument in the (0.1-2.4) keV
energy band is $< 0.9 \cdot 10^{-3} s^{-1}$ corresponding to an unabsorbed flux limit of
$F_{X} < 1.7 \cdot 10^{-14}$ \ergcm2s. This flux corresponds to an X-ray luminosity upper
limit of $L_{X} < 0.01 \cdot 10^{36}$ erg s$^{-1}$.

The main point we wish to make in the present analysis is that a complete multi-frequency analysis of the astrophysical DM signals
coming from Draco might carry much more information, and can be significantly more constraining, in terms of limits on DM WIMP models than, for instance, a study of the emissions in the gamma-ray frequency range alone.

As we show in the present analysis, available observational data, and the possible detection of WIMP annihilation signals from Draco by future instruments can be, in principle, of crucial relevance for the study of the nature of WIMP DM: the expected emission features
associated to DM annihilation secondary products are, in fact, the only radiation mechanisms which can be
expected in a system like a dSph, as originally envisioned by Colafrancesco
\cite{Colafrancesco2004,Colafrancesco2005}. Following our original suggestions, and
pursuing the systematic approach we outlined for the case of Coma (see CPU 2006), we present
here a detailed analysis and specific predictions for the WIMP DM annihilation signals expected from Draco
in a multi-wavelength strategy.

Specifically, we first derive the DM density profile of
Draco in a self-consistent $\Lambda$CDM scenario in Sect.~II. We then discuss the
gamma-ray emission produced in Draco from DM annihilation, assuming a set of model-independent
WIMP setups \cite{Colafrancesco:2005ji}. Gamma-ray emissions, and constraints, are studied in Sect.~III.
We then present in Sect.~IV the signals expected from Draco at all frequencies covered
by the radiation originating from the secondary products: synchrotron emission in the radio range,
Inverse Compton scattering of electrons and positrons produced by DM annihilation off CMB and starlight photons, and the associated SZ effect. We
also discuss, in Sect.~V the possible amplification of these signals by an
intervening black hole at the center of Draco. We present our conclusions in the
final Sect.~VI.

Throughout this paper, we refer to the concordance cosmological model suggested by WMAP 3yr.
\cite{Spergeletal2006}; namely, we assume that the present matter energy density is
$\Omega_m = 0.266$, that the Hubble constant in units of 100 km s$^{-1}$ Mpc$^{-1}$ is
$h=0.71$, that the present mean energy density in baryons is $\Omega_b  = 0.0233/h^2$,
with the only other significant extra matter term in cold dark matter $\Omega_{CDM} =
\Omega_m -\Omega_b$, that our Universe has a flat geometry and a cosmological constant
$\Lambda$, {\em i.e.} $\Omega_\Lambda = 1-\Omega_m$, and, finally, that the primordial
power spectrum is scale invariant and is normalized to the value $\sigma_8 = 0.772$.

\section{The dark matter density profile in Draco}
\label{sec:halo}

Modeling the distribution  of dark matter for dSph's is
not a straightforward task. The radial maps of the star velocity dispersions clearly
indicate that dSph are dark matter dominated systems. However, available observational
data do not provide enough information to unequivocally determine the shape and
concentration of the supporting dark matter density profiles (see e.g. the recent
analysis of Ref.~\cite{metal} for the case of Draco, under investigation here). Such
freedom is partially reduced restricting to $\Lambda$CDM inspired scenarios, as
appropriate for dark matter in the form of cold WIMP particles. Within this structure
formation picture, numerical N-body simulations of hierarchical clustering predict that
Milky Way size galaxies contain an extended population of  substructures, with masses
extending down to the free streaming scale for the CDM component (as small as $10^{-12} -
10^{-3} \msun$ in the case of neutralinos in supersymmetric models or in other WIMP
setups~\cite{Green:2005fa,Profumo:2006bv}), and surviving, at least in part, to tidal
disruption: dwarf satellites stand as peculiar objects, since they are the smallest ones
featuring a stellar counterpart, while mechanisms preventing star formation are
supposed to intervene for lighter objects (among scenarios supporting this
interpretation, see, e.g., \cite{nostar}). In case of isolated CDM halos, properties of
the dark matter density profile have been investigated in some detail through numerical
simulations: a universal shape and a correlation (on average) between the object mass and its
concentration are expected (more details will be given in the following section). The
picture is less clear for satellites, like Draco, standing well within the dark matter
potential well of the hosting halo. Tidal forces may have significantly remodeled the
internal structure of these objects, an effect which is likely to depend, e.g., upon the
merging history of each satellite. Based again on numerical simulations, significant
departures from the correlation between mass and concentration parameter observed for
isolated halos have been reported in the literature, as well as discrepant results
regarding whether the universal shape of the density profile is preserved \cite{pros} or
not\cite{cons} in the subhalos, after tides have acted and these systems have reached a
new equilibrium configuration.

\subsection{Mass models within the $\Lambda$CDM framework}

The main dynamical constraint we consider for mass models for the Draco dSph is the observed line-of-sight velocity dispersion of its stellar population. The 
underlying, necessary, assumption we shall make here is that the stellar component is in equilibrium, and hence
that the Jeans equation applies to this system; if this is the case, one finds that the projection
along the line of sight (l.o.s.) of the radial velocity dispersion of stars can be expressed in terms of
$M(r)$, the total (i.e. including all components) mass within the radius $r$ (\cite{BM,LM}):
\begin{eqnarray}
    \sigma_{\rm los}^2 (R) &=& \frac{2 G}{\Sigma(R)} \int_{R}^{\infty}
    \, {\rm d} r^\prime \ \nu (r^\prime) M(r^\prime) (r^\prime)^{2 \beta - 2}
    \int_{R}^{r^\prime}  \,{\rm d} r \left( 1-\beta \frac{R^2}{r^2} \right)
    \frac{r^{-2 \beta +1}}{\sqrt{r^2 - R^2}} \ ,
    \label{eq:los}
\end{eqnarray}
where $\nu(r)$ is the density profile of the stellar population and $\Sigma(R)$ represents its surface
density at the projected radius $R$. In the derivation of Eq.~(\ref{eq:los}), we  have assumed that
the anisotropy parameter $\beta$ is constant over radius; in terms of the radial and tangential
velocity dispersion, respectively $\sigma_r$ and $\sigma_\theta$,
$\beta = 1- \sigma_\theta^2 / \sigma_r^2$: $\beta=1$ denotes the case of purely radial orbits,
$\beta=0$ that of a system with isotropic velocity dispersion, while $\beta \rightarrow - \infty$
labels circular orbits. As we will see shortly, the anisotropy parameter is important since we
recover in our analysis the well known degeneracy between the reconstructed mass profile and
the assumed degree of stellar anisotropy.

\begin{figure*}[!t]
\begin{center}
\includegraphics[scale=0.55]{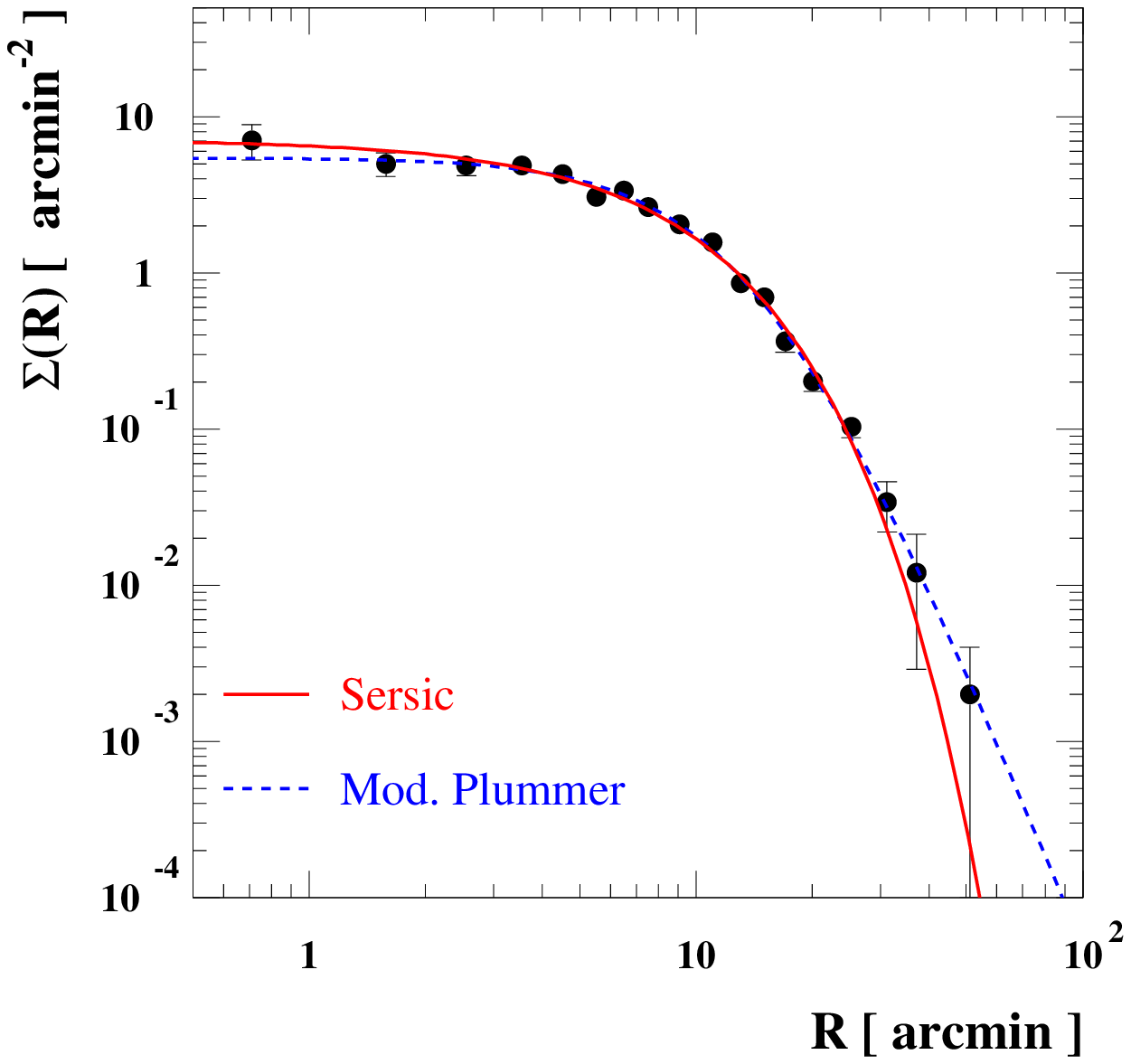}
\quad\includegraphics[scale=0.55]{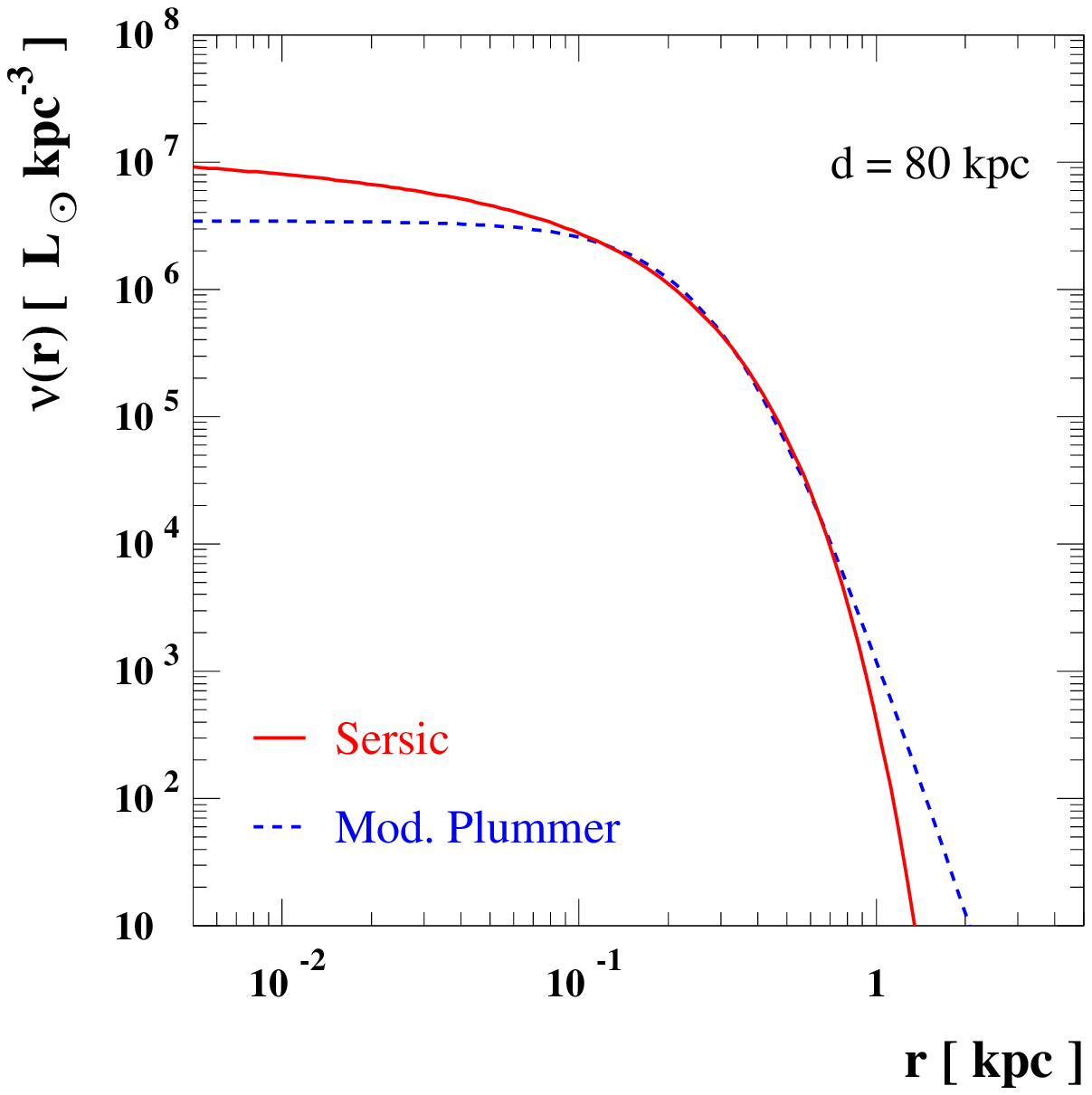}\\
\end{center}
\caption{Left panel: radial profile for the surface brightness distribution of stars in Draco;
data are from Odenkirchen et al.~\protect{\cite{odenetal}}, while fits are with a Sersic profile or
a modified Plummer model. Right panel: the corresponding luminosity density profiles.}
\label{fig:surf}
\end{figure*}

Extensive photometric studies are available for the Draco dwarf; we refer to the analysis in
Odenkirchen et al.~\cite{odenetal} relying on multicolor data from the SDSS (sample and
foreground determination labeled S2 in that analysis) and reproduce the result for the radial
profile of the surface brightness in Fig.~\ref{fig:surf} (left panel). We also show two
alternative fits of the data:  one option is the generalized exponential profile proposed by
Sersic~\cite{sersic} and implemented in the case of Draco also by Lokas, Mamon \& Prada~\cite{LMP}:
\begin{equation}
   \Sigma(R) = \Sigma_0 \exp [-(R/R_{\rm S})^{1/m}]\,,
   \label{eq:sersic}
\end{equation}
choosing the parameter $1/m = 1.2$, and fitting the scale radius $R_{\rm S}$ and central
surface brightness $\Sigma_0$ to the data (the best fit procedure gives $R_{\rm S}=7\farcm3$).
As second possibility, we follow Mashchenko et al.~\cite{metal} and consider  a modified
Plummer model:
\begin{equation}
   \Sigma(R) = \Sigma_0 \left[ 1 + (R/R_{\rm P})^2\right]^{-(\alpha-1)/2},
   \label{eq:modplummer}
\end{equation}
setting the exponent $\alpha=7$, and then fitting the value for the scale radius
($R_{\rm P}=14\farcm6$). For each of the two $\Sigma(R)$, the luminosity density profile
$\nu(r)$ is obtained by inverting the definition of surface brightness with the Abel integral
formula, i.e. implementing the de-projection:
\begin{equation}
    \nu(r) = - \frac{1}{\pi} \int_r^\infty dR
    \frac{1}{\sqrt{R^2 - r^2}}\, \frac{d\Sigma}{dR}.
\end{equation}
The inversion is performed numerically for the Sersic profile, while it can be done analytically
for the modified Plummer model; results are shown in Fig.~\ref{fig:surf} (right panel) and
one can see that the mild differences in the surface brightness are only marginally
amplified in the luminosity density profiles.
Here we are referring to luminosities in the V-band and, following again~\cite{LMP},  we
have adopted for the distance of Draco the value 80~kpc~\cite{aparicio}, or, equivalently, a distance
modulus of 19.5~\cite{cioni}, standing in between (and in agreement at 1~$\sigma$) the
other recent estimate of $75.8 \pm 0.7 \pm 5.4$~kpc from Ref.~\cite{bonanos} and the value
of $82 \pm 6$~kpc from the compilation of Mateo~\cite{mateo}. To add the stellar component
in the total mass term $M(r)$ in Eq.~(\ref{eq:los}), we need an estimate for the stellar
mass-to-light ratio in the V-band; we mainly refer to one of the largest values quoted in the literature,
$\Upsilon_V = 3 M_{\sun}/L_{\sun}$, including in it a possible subdominant gas component.

The ansatz we implement for the dark matter component is that of a spherical distribution sketched
by a radial density profile:
\begin{equation}
   \rho (r)=\rho^{\prime} g(r/a)\,;
\label{eq:profi}
\end{equation}
given in terms of the function $g(x)$ and of two parameters, i.e. a scale radius $a$ and a
normalization factor $\rho^{\prime}$. This is in analogy with the usual description of dark matter
halos from results of numerical N-body simulations in terms of a universal density profile;
we take as guideline for our mass models the form originally proposed by  Navarro,
Frenk \& White~\cite{NFW}:
\begin{equation}
   g_{NFW}(x) = \frac{1}{x \, (1+x)^2}
\label{eq:nfw}
\end{equation}
and a shape slightly more  singular towards the center proposed by Diemand et al.~\cite{d05}
(hereafter labeled as D05 profile):
\begin{equation}
   g_{D05}(x) = \frac{1}{x^{\gamma} (1+x)^{3-\gamma}}
  \;\;\;\;{\rm with} \;\;\;\; \gamma \simeq 1.2\;.
\label{eq:d05}
\end{equation}
As a further option, we consider the Burkert profile (\cite{burkert}):
\begin{equation}
   g_{B}(x) = \frac{1}{(1+x)\,(1+x^2)}\;,
\label{eq:burk}
\end{equation}
i.e. a model with a large core, in agreement with the gentle rise in the inner part of the rotation
curves occurring in a vast class of galaxies, including dwarfs~\cite{burkert,salucci}.
Mechanisms of gravitational heating of the dark matter by baryonic
components during or after the baryon infall have been advocated
to reconcile these observations with the central density cusps of the profiles  introduced above~\cite{WK,el-zant,MCW};
these models are still contrived and it is probably premature to say whether in the case of Draco
a cored or cuspy halo is expected.

Since we shall extrapolate the dark matter mass profile well
beyond the radial size of the stellar component, we need a description of the regime where the profile
gets sensibly reshaped by tidal interactions with the dark matter halo of the Milky Way.
We compute the tidal radius $r_{\rm tid}$ in the impulse approximation, as appropriate for
extended objects~\cite{tormen,hayashi}:
\begin{equation}
  \frac{M(r_{\rm tid})}{r_{\rm tid}^3} = \left. \left[ 2-\frac{r}{M_{MW}(r)}
  \frac{{\partial M}_{MW}}{\partial r}\right]
  \frac{M_{MW}(r)}{r^3}\right|_{r=r_p-r_{\rm tid}}\,, gf  \label{eq:rtid}
\end{equation}
with $M(r_{\rm tid})$ the mass of Draco within the tidal radius, and $M_{MW}(r)$ the mass
of the Milky Way within the galactocentric distance $r$; the expression on the right hand
side is computed for the orbital radius of Draco $r_p$ at its latest pericenter  passage.

Mass models for Draco are generated as follows: for a given functional form for the profile
and for any given pairs of the parameters $\rho^{\prime}$ and $a$, the density profile is shifted
into the form~\cite{kaza}:
\begin{equation}
  \rho (r) \rightarrow \rho(r) \exp(-r/r_{\rm tid})
\label{eq:proftidal}
\end{equation}
with $r_{\rm tid}$ determined  from Eq.~(\ref{eq:rtid}), assuming for the Milky Way a virial mass
equal to $10^{12}\,\msun$ and a NFW profile with concentration parameter equal to
13~\cite{zks}; $r_p$ will be taken, as a first test case,
equal to 20~kpc, which is about the minimum pericenter
radius below which tidal effects would be visible in  the stellar component as well~\cite{metal},
and which gives the most conservative estimate for the dark matter mass in Draco.
We are then ready to implement Eq.~(\ref{eq:los}) and compare against data.

\begin{figure*}[!t]
\begin{center}
\includegraphics[scale=0.55]{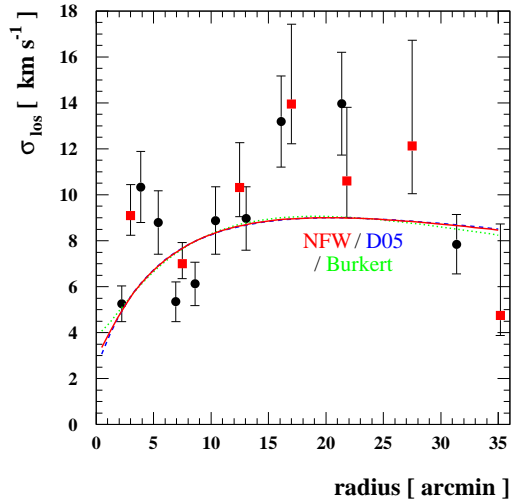}\\
\end{center}
\caption{Line of sight velocity dispersion for the three best-fit-models considered
in the paper. Data are from Munoz et al. 2005 (black filled circles) and from
Wilkinson et al. (red filled squares).
} \label{fig:veldisp}
\end{figure*}

Munoz et al.\cite{munozetal} have recently made a novel compilation of l.o.s. star velocity
dispersions in Draco, containing 208 stars; they show results implementing several binning criteria,
among which we resort to the one with the largest number of stars per bin (21 stars per bin),
which is the least susceptible to statistical fluctuations.
Using essentially the same data sample, but a different binning, Wilkinson et al.~\cite{wilkinsonetal}
find a sharp drop in the velocity  dispersion corresponding to the bin at the largest circular radius,
a feature that does not emerge in the analysis of Munoz et al. On the other hand,
Lokas, Mamon \& Prada~\cite{LMP} question whether this sample should be further cleaned
from outliers, i.e. stars that may not actually be gravitationally bound to Draco.
We will compare separately with the data set from Munoz et al., i.e. in 10 bins out to a circular
radius of slightly larger than $30^\prime$, and the one from Wilkinson et al., i.e. 7 bins out to a circular
radius of about $35^\prime$, see Fig.~\ref{fig:veldisp}.
For any mass model we consider the reduced $\chi^2$ variable:
\begin{equation}
  \chi^2_{red}=
  \frac{1}{N_{\rm bins}} \sum_{j =1}^{N_{\rm bins}} \frac{\left( \sigma_{\rm los}(R_j)-\sigma_{\rm los}^j
  \right)^2}{(\Delta \sigma_{\rm los}^j)^2}\,.
\label{eq:chisq}
\end{equation}
$\chi^2_{red}$ is very sensitive to the value of the overall normalization parameter, moderately
sensitive to $\beta$, while it is less sensitive to the length scale $a$. In Fig.~\ref{fig:veldisp} we
show the line of sight velocity dispersion projected along the l.o.s., comparing to the
Munoz et al. dataset and assuming the star distribution according to the Sersic profile. The
best fit models for the three dark matter density profiles we are focusing on are set as follows:
{\sl i)} a NFW profile with $a= 1$~kpc, $\rho^\prime = 3.7 \, 10^7 \msun$~kpc$^{-3}$,
$r_{tidal} = 1.7$~kpc and $\beta= -3.7$; {\sl ii)} a Burkert  profile with $a= 0.5$~kpc,
$\rho^\prime = 2.1 \, 10^8 \msun$~kpc$^{-3}$, $r_{tidal} = 2.0$~kpc  and $\beta= -1.0$;
{\sl iii)} a D05 profile with $a= 1$~kpc, $\rho^\prime = 2.54 \, 10^7 \msun$~kpc$^{-3}$,
$r_{tidal} = 1.5$~kpc  and $\beta= -6.3$.  Clearly, the dataset does not allow for a discrimination
among the three models.

\begin{figure*}[!t]
\begin{center}
\includegraphics[scale=0.55]{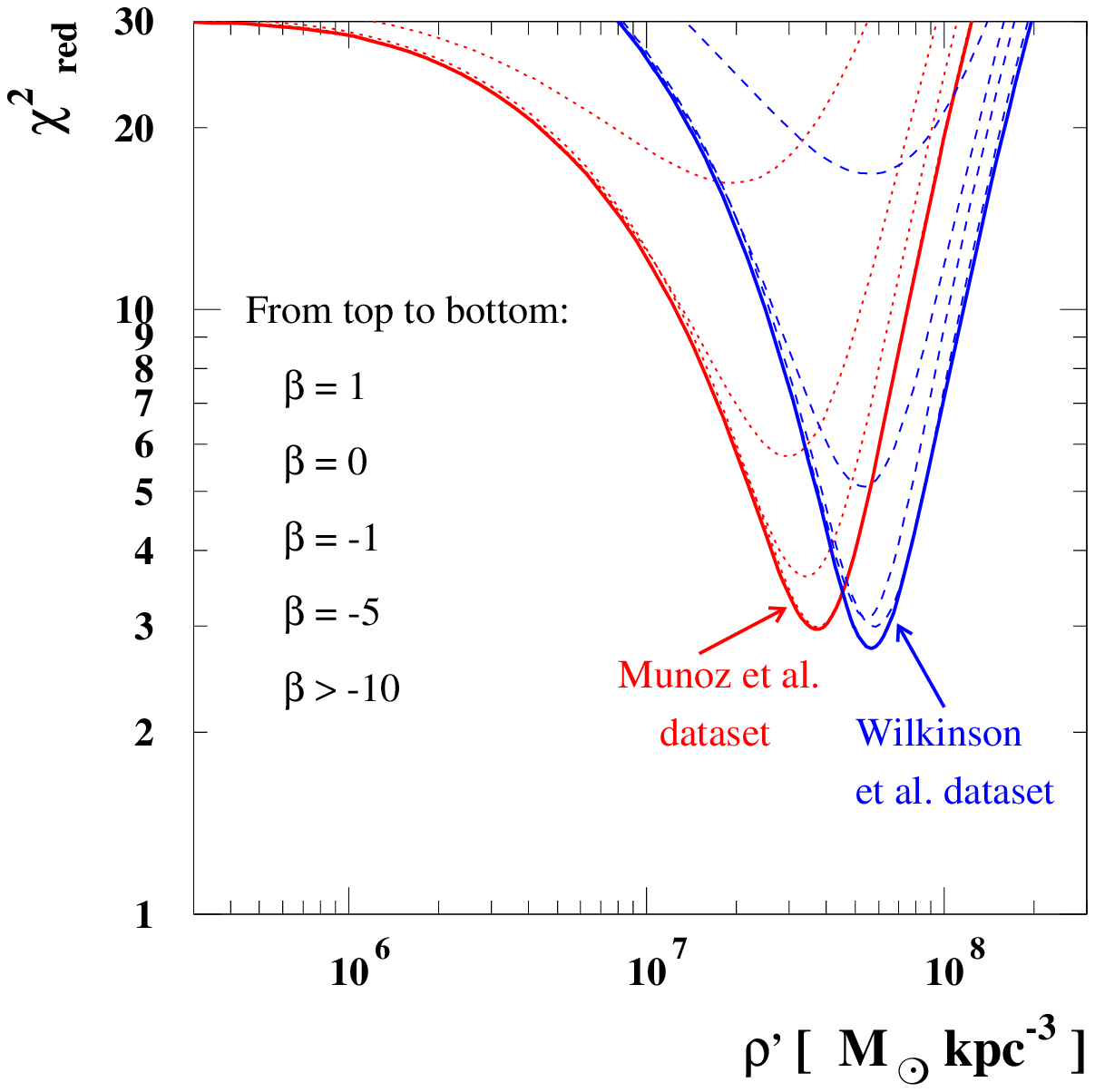}
\quad\includegraphics[scale=0.55]{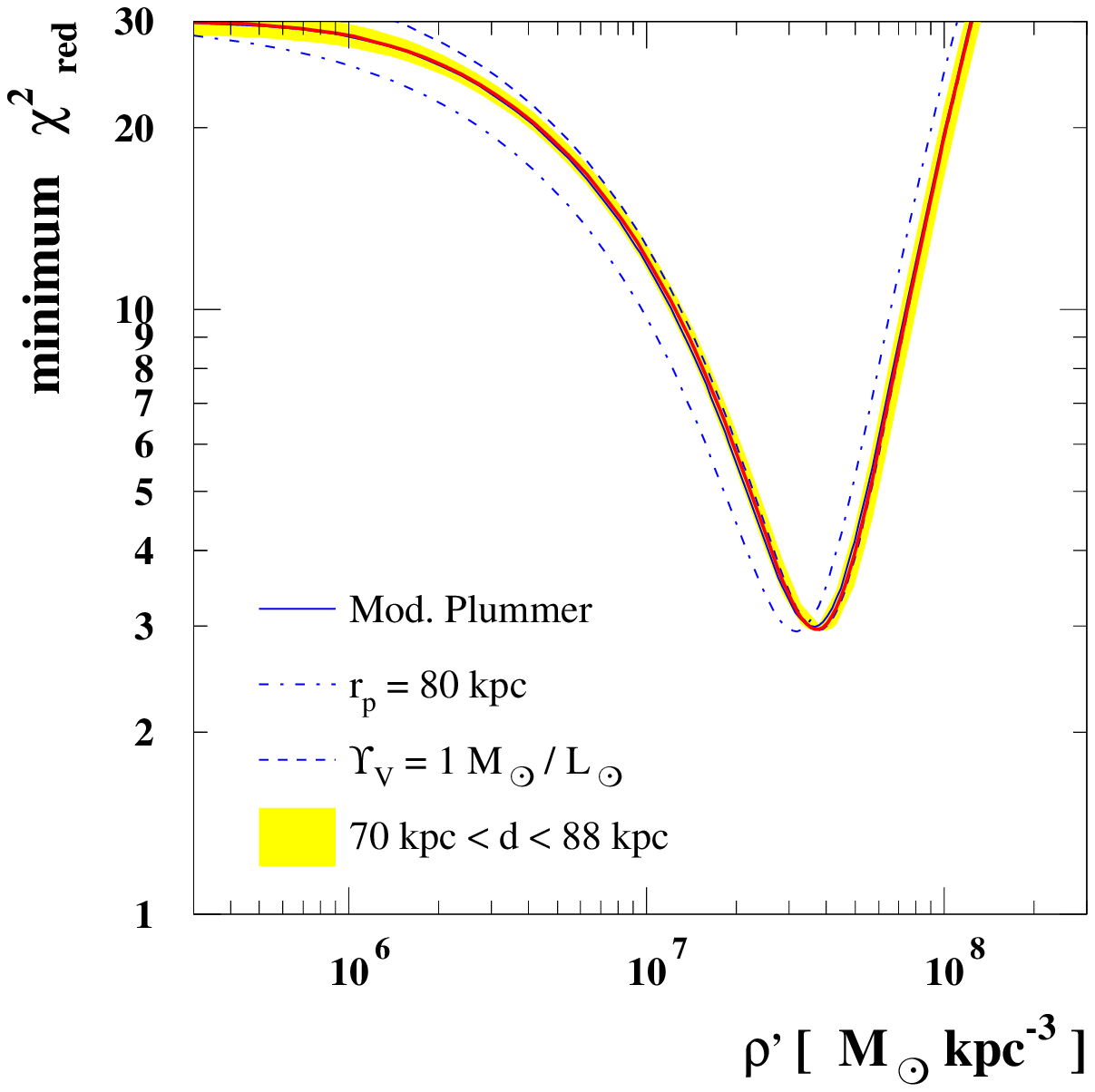}\\
\end{center}
\caption{We consider the NFW profile as reference halo model, fix the length scale
parameter to 1~kpc and plot the reduced $\chi^2$ as a function of the density normalization
parameter $\rho^\prime$. In the left panel we plot the reduced $\chi^2$ obtained either for
the Munoz et al. or Wilkinson et al. data sets, for a few selected values of the anisotropy
parameter $\beta$ or choosing the parameter $\beta$ (in the interval $-10 < \beta < 1$)
which at a given $\rho^\prime$ gives the smallest reduced $\chi^2$;
other underlying assumptions (default model) are: radial star profile described  by a Sersic profile,
distance at latest pericenter passage $r_p = 20$~kpc, mass-to-light ratio of 3~
$\msun / L_{\odot}$ and distance of Draco $d =  80$~kpc. In the right panel we show
that none of these latter assumptions are crucial: we plot as a function of $\rho^\prime$
the minimum reduced $\chi^2$ for $\beta$ between -10 and 1, in case of radial star profile
according to the modified Plummer model, $r_p$ increased to 80~kpc, mass-to-light ratio
decreased to 1~$\msun / L_{\odot}$ and distance of Draco varied within a generous
range of values.
 }
\label{fig:halonfw}
\end{figure*}

In Fig.~\ref{fig:halonfw} we illustrate the sensitivity of the fit to some of the parameters
introduced in our model, taking the NFW profile as reference case, and $a= 1$~kpc as in the
best fit model: the minimum of $\chi^2$ is well defined with respect to $\rho^\prime$ and has
a marginal shift when comparing to the data as in the binning of Wilkinson et al.; had
we followed the suggestion of Ref.~\cite{LMP} to take out of the sample some of the stars that
appear as outlier,  the minimum reduced $\chi^2$ would get below 1, but its position on the
the $\rho^\prime$ axis would not change appreciably. Also shown is the dependence
of the result upon the anisotropy parameter $\beta$: for the NFW profile, the case of radial orbits
is disfavored, while models with circular anisotropy give better fits. In the right panel of
Fig.~\ref{fig:halonfw} we show instead that none of our additional  assumptions has a significant
impact on the velocity dispersion fit. In particular, there is a marginal effect when considering
an alternative fit to the stellar profile, or when varying the assumed value for the distance of the
Draco within the ranges of estimates quoted in the literature, or when decreasing the mass-to-light
ratio of the stellar component to significantly smaller values.
Also secondary, but slightly larger, is the effect of assuming that the current position of
Draco is also the smallest galactocentric distance reached so far in its orbital motion,
and hence it is the relevant radius to estimate the effects of tidal stripping (in this case, tidal radii become
much larger than the scale radius for the stellar component).

\begin{figure*}[!t]
\begin{center}
\includegraphics[scale=0.55]{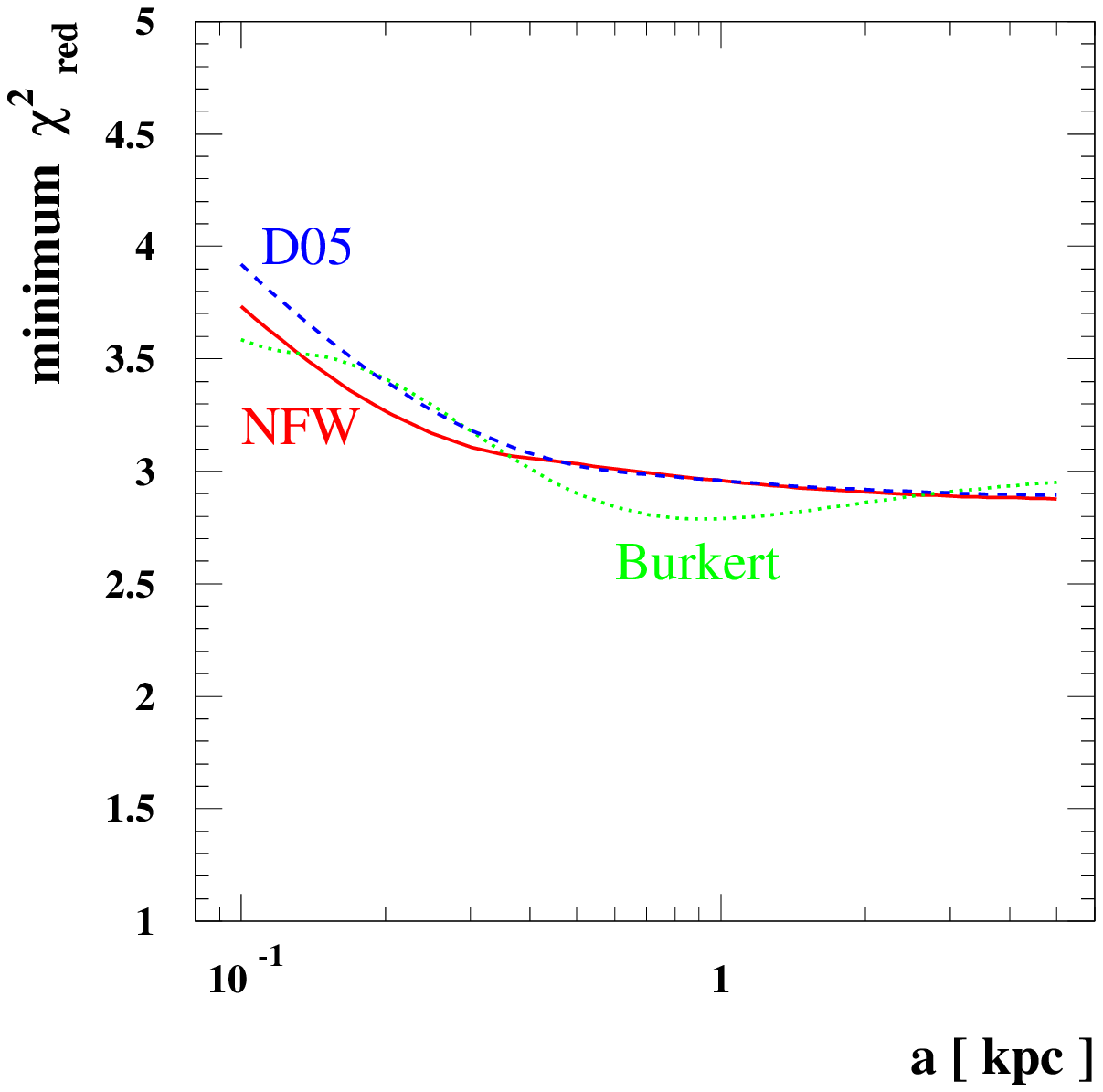}
\quad\includegraphics[scale=0.55]{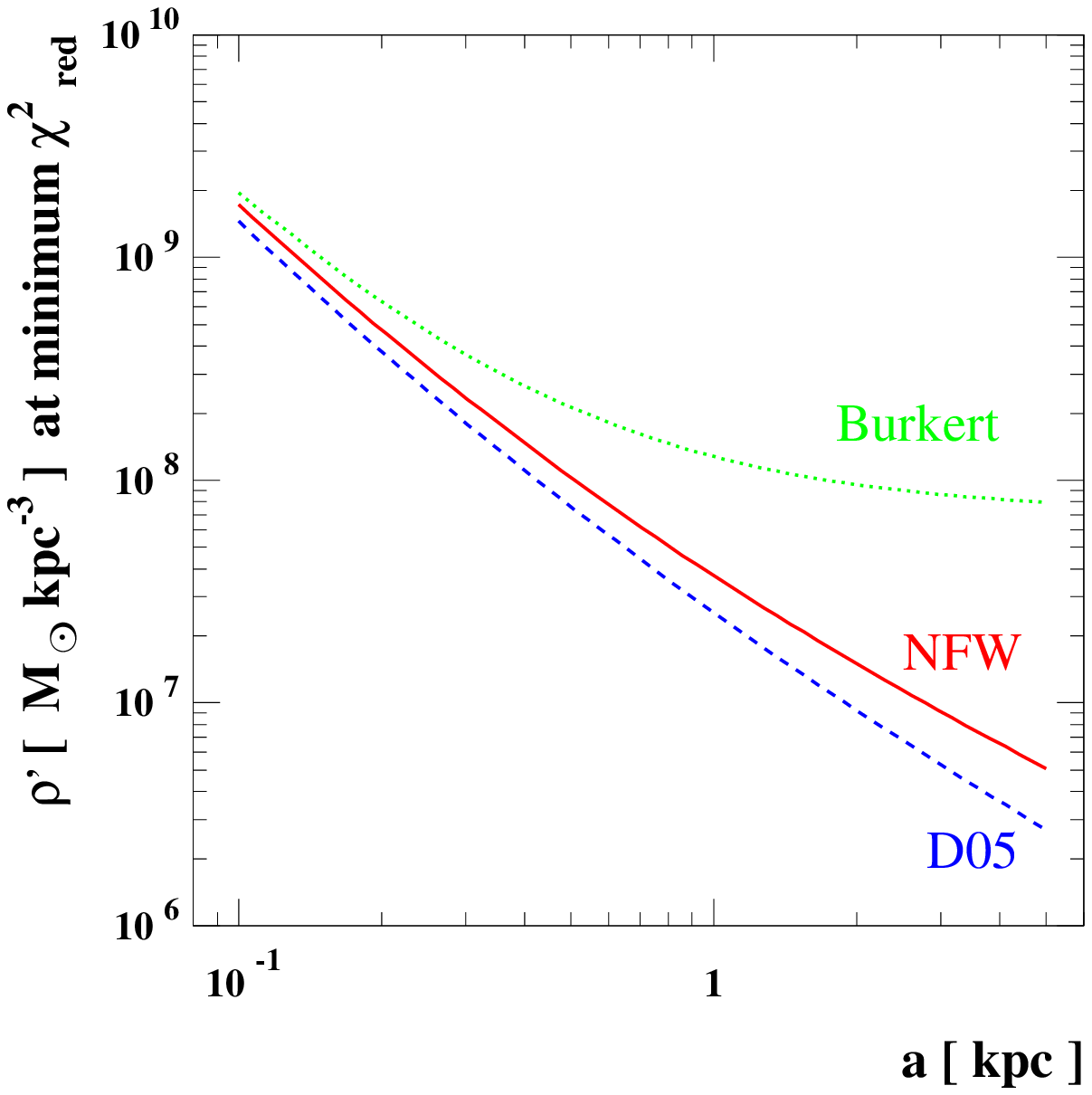}\\
\end{center}
\caption{Left panel: for a given value of the length scale parameter $a$, we plot the
minimum $\chi^2_{red}$ obtained by varying the parameter $\rho^\prime$ and the
anisotropy parameter $\beta$ between -10 and 1; we consider the three dark matter
halo profiles introduced in the text,  compare against the Munoz et al. dataset and refer
to our default model for the other parameters. Right panel: for a given value of
the length scale parameter $a$, value of $\rho^\prime$ corresponding to the minimum
$\chi^2_{red}$ displayed in the left panel for the three dark matter halo profiles.
} \label{fig:halo1}
\end{figure*}

In Fig.~\ref{fig:halo1} we show the minimum value of the reduced $\chi^2$, obtained
taking the density normalization $\rho^\prime$ and the stellar anisotropy $\beta$
as free parameters, for the three dark matter density profiles and as a function of the
scale factor $a$: as clearly emerging from the Figure, the dataset does not allow for a clear discrimination
in the parameter $a$, but there is, rather, a close correlation between length scale and
density normalization parameter. In the right panel of Fig.~\ref{fig:halo1} we plot the
value of $\rho^\prime$ corresponding to the model with  minimum $\chi^2$ and a given scale factor $a$;
note the huge span in the range of values of the logarithmic vertical scale.

\begin{figure*}[!t]
\begin{center}
\includegraphics[scale=0.55]{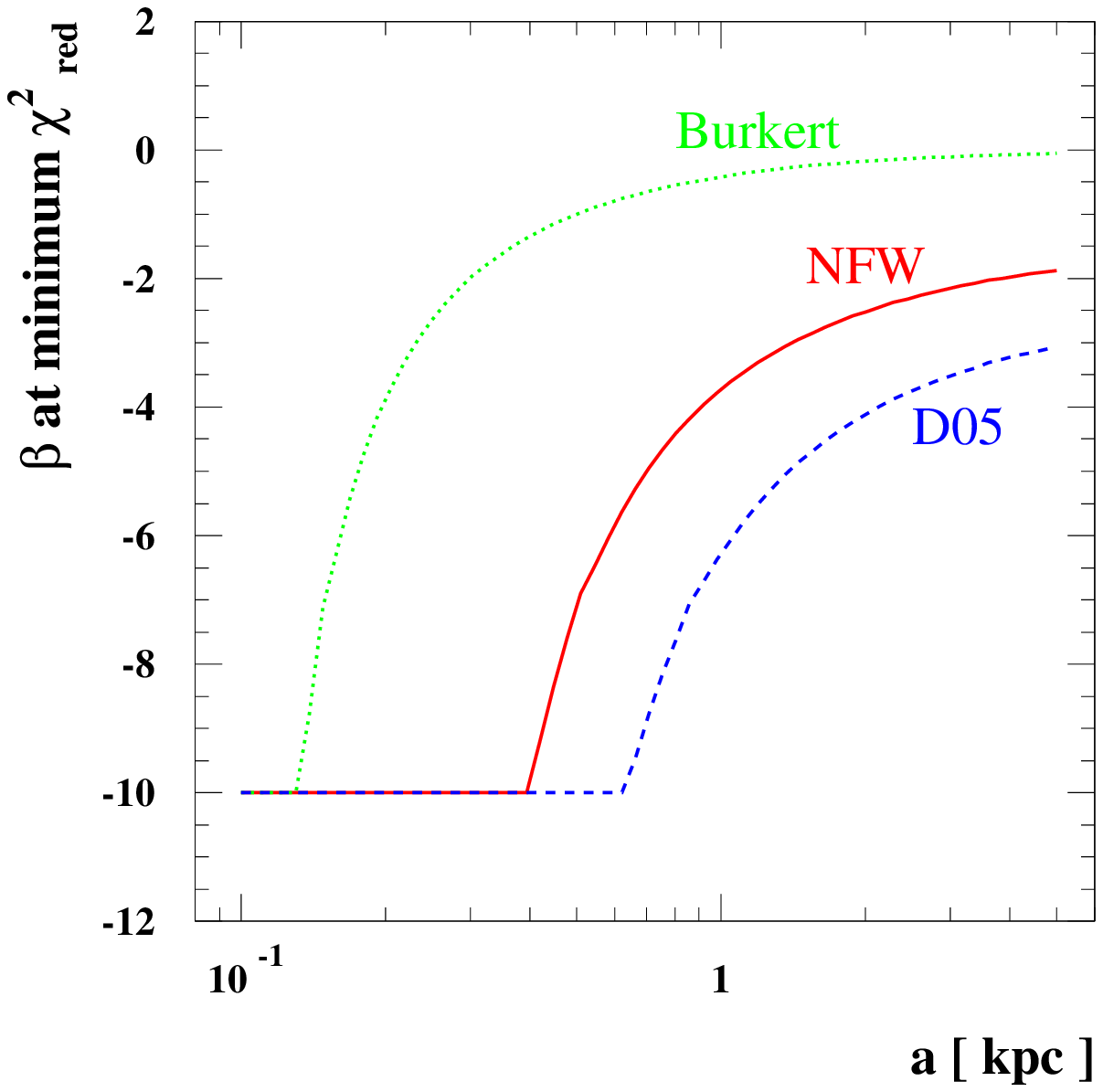}
\quad\includegraphics[scale=0.55]{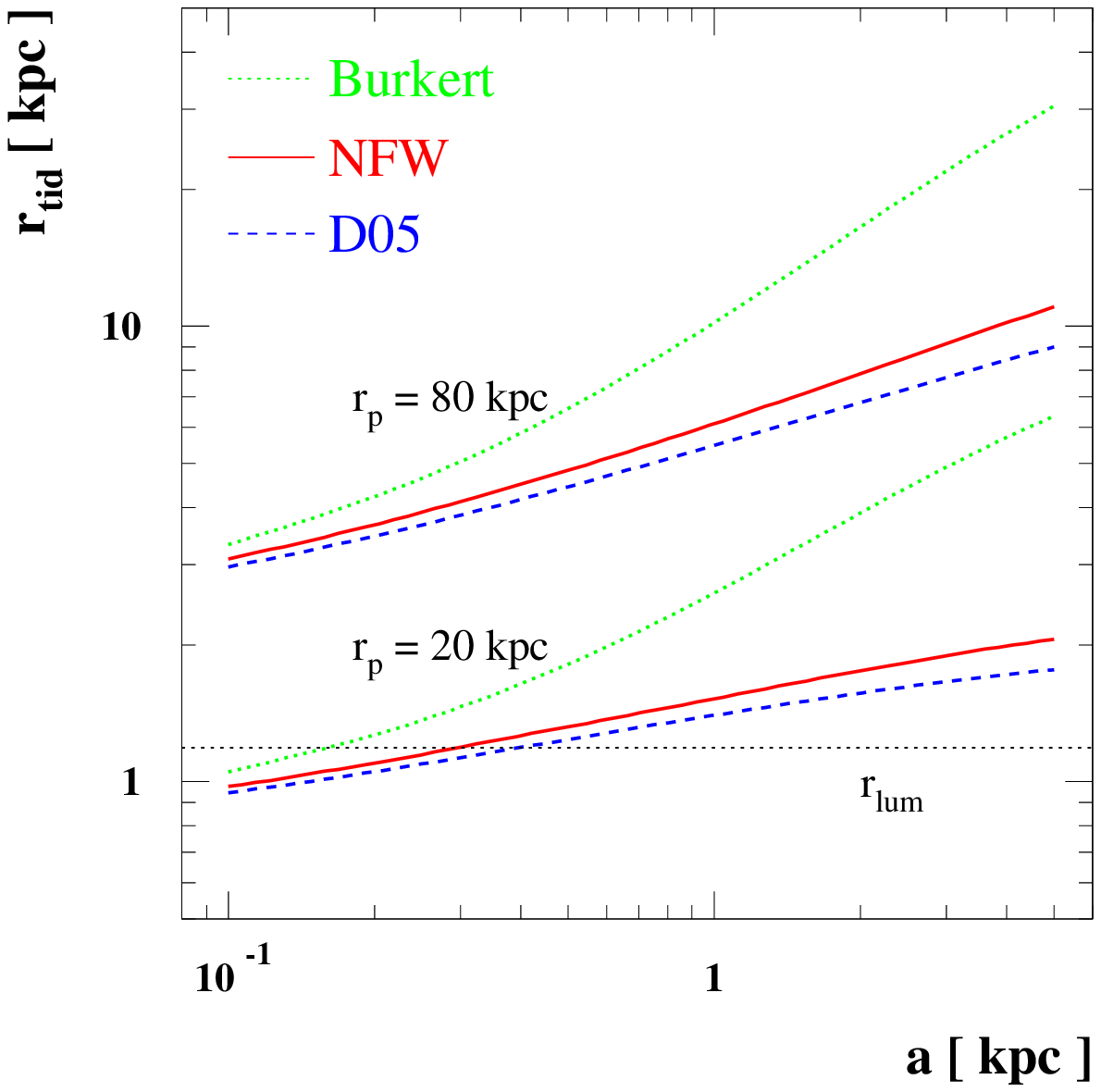}\\
\end{center}
\caption{Values of $\beta$ (left panel) and of the tidal radius (right panel, lower curves
corresponding to a radius at latest pericenter passage equal to 20~kpc) defining the
model with minimum $\chi^2_{red}$ at a given $a$ displayed in the left panel of
Fig.~\protect{\ref{fig:halo1}}, for the three dark matter density profiles. We also show in the right panel the value of the tidal radius setting the radius at latest pericenter passage to 80~kpc, and keeping other parameters unchanged; values for  $\chi^2_{red}$
change marginally, i.e. form the point of view of fitting the data the two cases are
equivalent, but obviously all $r_{tid}$ increase significantly (for reference we
plot with a dotted horizontal line the radial size of stellar component).
} \label{fig:halo2}
\end{figure*}

In Fig.~\ref{fig:halo2} we
show the tidal radii as determined assuming for the radius at the last pericenter passage
20~kpc or 80~kpc (right panel), and values of $\beta$ (left panel) set as in the best fit models;
shallower, or less concentrated, profiles give equivalent fits to the data if
the degree in circular anisotropy is decreased
($\beta = -10$ is the minimum value we are scanning on;
isotropic, $\beta = 0$, models are favored for the cored Burkert profile in the case of moderate
to large values for the scale factor).

\subsection{Connections  to the structure formation picture}

\begin{figure*}[!t]
\begin{center}
\includegraphics[scale=0.55]{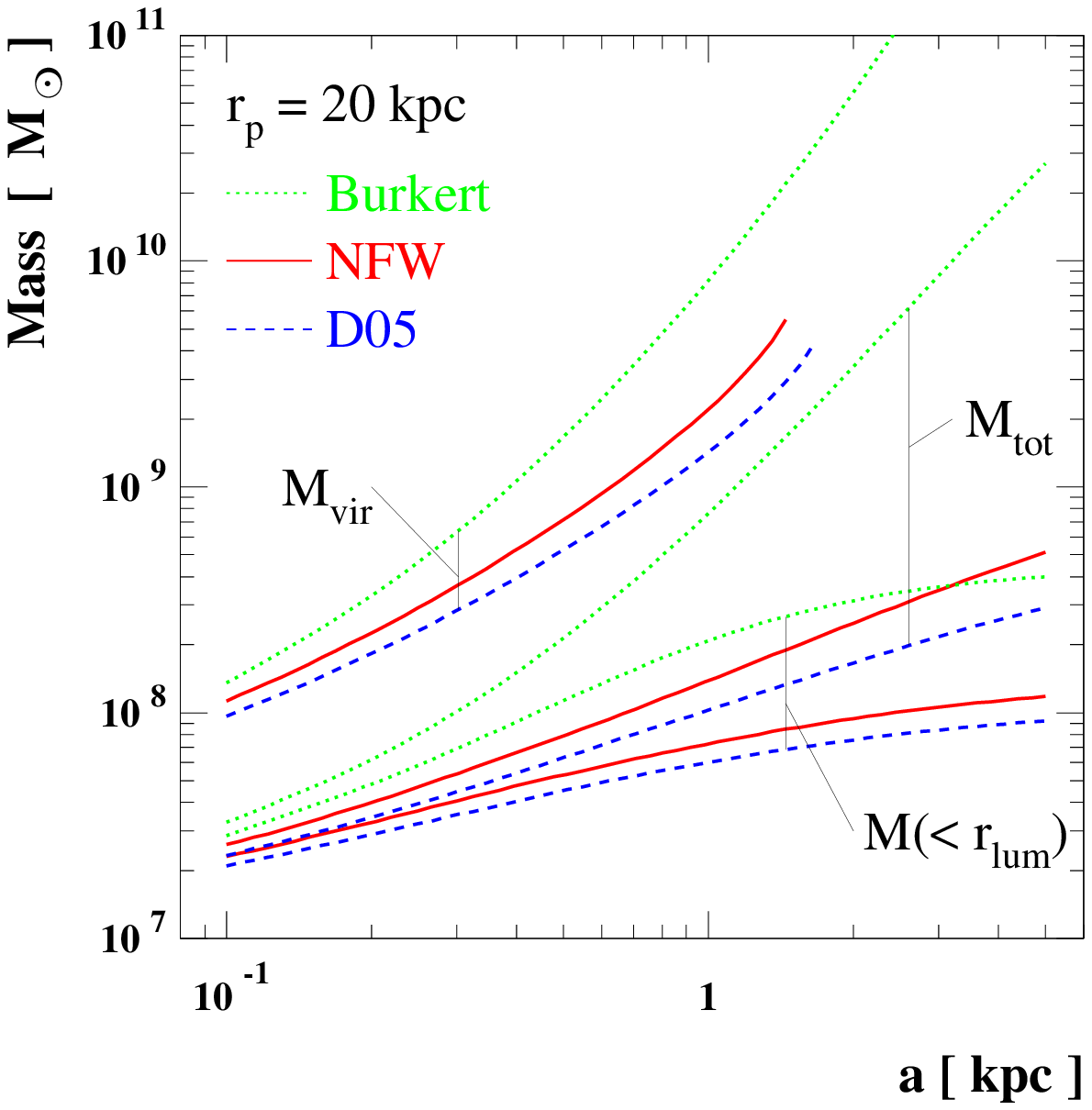}
\quad\includegraphics[scale=0.55]{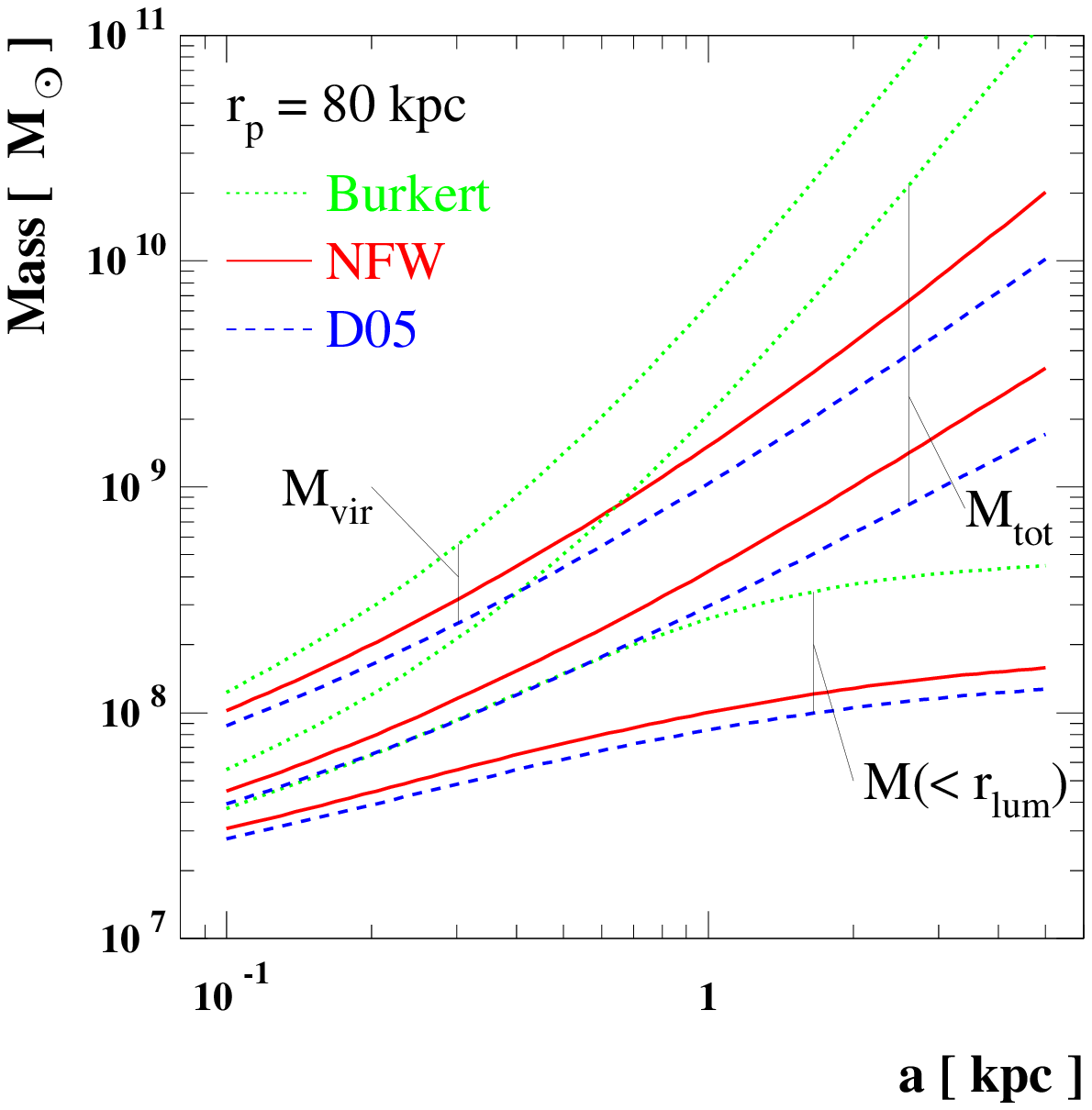}\\
\end{center}
\caption{We display, for the minimum $\chi^2_{red}$ models illustrated in
Figs.~\protect{\ref{fig:halo1}} and \protect{\ref{fig:halo2}}, values of the dark matter
halo mass within the radial size of stellar component (lower curves) and of the
total mass in the dark matter component for the calculated tidal
radius (medium curves), assuming a pericenter radius of 20~kpc (left panel)
and 80~kpc (right panel). We have also performed an extrapolation to estimate
the initial virial mass  of Draco, i.e. the mass associated to it before sinking deep
into the potential well of the Milky Way and the loss of a large fraction of such initial
mass due to tidal effects.
} \label{fig:masses}
\end{figure*}

\begin{figure*}[!t]
\begin{center}
\includegraphics[scale=0.55]{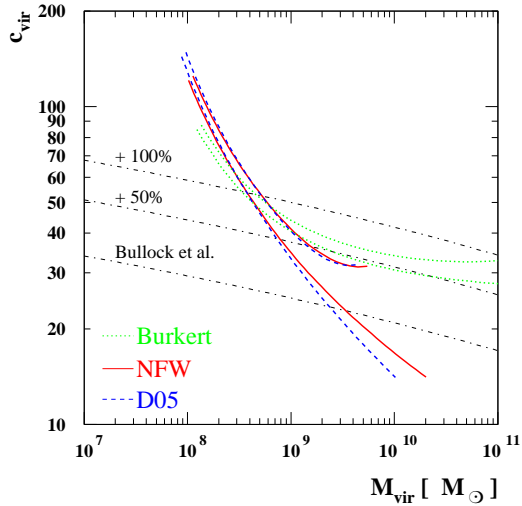}
\end{center}
\caption{We display, for the minimum $\chi^2_{red}$ models illustrated in
Figs.~\protect{\ref{fig:halo1}} and \protect{\ref{fig:halo2}}, the extrapolated values of
the virial mass and concentration parameter according to the prescription for the
response to tidal interactions introduced in \cite{hayashi} and fitted to numerical
simulations. Also shown is the extrapolation with the Bullock et al. prescription of the
correlation  $M_{vir} - c_{vir}$ for isolated halos, and assuming a 50\% or a 100\%
increase in concentration in the case of subhalos. } \label{fig:mvir}
\end{figure*}

The possibility of discriminating among dark matter halo models increases when we take
into account results from N-body simulations of structure formation. To make this step we
need, however, to rely on a series of extrapolations. The first is to try to map the fit
we made for a tidally disrupted object, well within the Milky Way potential well, to the
configuration of a virialized system, unaffected by tides, of the kind described, on
statistical grounds, by results of simulations. We refer to the prescription derived from
numerical studies in \cite{hayashi}: let the density profile prior to tidal interactions
be in the form:
\begin{equation}
   \rho_{{\rm no \; tides}}(r)= \rho_s \, g(r/r_s)\,;
\end{equation}
Suppose, then, that tidal interactions change it into the form:
\begin{equation}
   \rho(r) = f_t \exp(-r/r_{\rm tid}) \; \rho_{{\rm no \; tides}}(r),
\label{eq:proftidalsf}
\end{equation}
assuming that the length scale parameter $a$ in the final profile is equal to the
initial scale factor $r_s$. Comparing  the form of Eq.~(\ref{eq:proftidal}) to the
one we used in the fit to the stellar velocity dispersion, i.e. Eq.~(\ref{eq:proftidalsf}),
we find $\rho^{\prime} = f_t \,\rho_s$. The parameter $f_t$ is a dimensionless
measure of the reduction in central density due to tidal effects; simulations
indicate that the latter is correlated to the mass fraction of the satellite bound to
the object after the effect of tides, $m_{\rm bnd}$, through the expression~\cite{hayashi}:
\begin{equation}
   \log{f_t} = - 0.007 +0.35 \log{m_{\rm bnd}} + 0.39 \left( \log{m_{\rm bnd}} \right)^2
   + 0.23  \left( \log{m_{\rm bnd}} \right)^3
\label{eq:ft}
\end{equation}
(we will assume this phenomenological fitting formula to be valid for $m_{\rm bnd}$
larger than about 5\%). According to this scheme, we can uniquely assign to any best
fit model with given $\rho^{\prime}$ and $a$ (for an assumed pericenter radius through which
$,r_{\rm tid}$ is determined) the corresponding $\rho_s$ and $r_s$, or equivalently a value
for the initial virial mass of the object $M_{vir}$ and its concentration parameter $c_{vir}$,
defined as $c_{vir} ={R_{vir}}/{r_{-2}}$. In this last step we introduced the virial radius
$R_{vir}$, defined as the radius within which the mean density of the halo is equal to the virial
overdensity $\Delta_{vir}$ ($\simeq 340$ at z=0) times the mean background density,
and the radius $r_{-2}$ where the effective logarithmic slope of the density profile
is $-2$ ($r_{-2}$ is
equal to $a$ for the NFW profile, $0.8\, a$ for the D05 and about $1.5\, a$ for the Burkert
profile). In Fig.~\ref{fig:masses} we plot $M_{vir}$ for the best fit models displayed in
Figs.~\ref{fig:halo1} and \ref{fig:halo2}; for comparison, we also show the total halo mass
bound to Draco after tidal stripping, and the dark matter mass within the spherical shell
defined by the radius of the stellar component, i.e. $r_{\rm lum} = 51^\prime$.
We have referred to the two extreme choice of pericenter radii, i.e. 20~kpc and 80~kpc;
the procedure seems fairly consistent since in the two cases we get very close
values for $M_{vir}$ (in the case of a small pericenter radius and the NFW or D05 profile,
at large $a$ the fraction of mass loss becomes very large and extrapolations according
to Eq.~(\ref{eq:ft}) becomes unreliable, so values of $M_{vir}$ are not displayed).
In Fig.~\ref{fig:mvir} we plot, for the same best fit models, the concentration parameter
versus virial mass; we also show the $M_{vir} - c_{vir}$ correlation as extrapolated,
for the currently preferred cosmological setup~\cite{WMAP}, from the toy model of
Bullock et al. ~\cite{Bullock}, which is tuned to reproduce the scaling found
in numerical simulations for isolated halos. As far as substructures are concerned, concentrations are
expected to be systematically larger, since substructures form, on average, in a denser environment
 with respect to isolated halos; for illustrative purposes
only, we show the $M_{vir} - c_{vir}$ scaling in the case of a 50\% and a 100\% increase
in concentration.

We have already stressed a few times that our analysis is heavily relying on
extrapolations, so no firm conclusion can be derived; nevertheless, our results seem
to indicate that we should prefer models  with an intermediate $M_{vir}$, say $10^9 \msun$,
corresponding to $a$ of the order of 1~kpc for the NFW and D05 profiles  and about
$0.5$~kpc for the Burkert profile (such cases are those that we have been chosen as
reference models in Fig.~\ref{fig:veldisp}) and that the range of length scale values allowed
in Figs.~\ref{fig:halo1} and \ref{fig:halo2} is probably a very generous one, with values
at the lower and upper ends which should be most likely dropped. The range of models
we are indicating here as preferred by the NFW profile is analogous to the one suggested
in Ref.~\cite{metal}, although the two approaches differ. In particular we will
not implement here a constraint from the age of Draco stellar population which is used as
guideline in~\cite{metal}: to do that we would need to build a subhalo mass function for
the Milky Way matching the observed satellite pattern, and to model star formation within
subhalos, two steps which are not very well understood and on which the degree of
extrapolation would be inevitably much more drastic than what we have accepted so far.

\section{The gamma-ray signal from WIMP annihilations in Draco}

WIMPs have a small but finite probability to annihilate in pairs, giving rise to potentially
observable standard model yields. Two ingredients intervene in fixing source functions:
on the one hand, the annihilation cross section, branching ratios and spectral distributions
for the yields are specified in any given particle physics scenario embedding the WIMP;
on the other hand, source functions scale with the number density of WIMP pairs, i.e. in the case
we are considering here, they are proportional to the square of the dark matter mass density
in Draco. Since photons in the energy range we are interested to, i.e smaller
than few TeV, propagate through the interstellar medium without being absorbed,
predictions  for the induced gamma-ray fluxes are straightforward and simply involve an
integral of the source along the line of sight; the expression for the flux per unit energy
and solid angle, is usually casted in the form:
\begin{equation}
   \phi_{\gamma}(E_\gamma,\Theta, \Delta\Omega)= \frac{(\sigma v)}{8\,\pi\,M_{\chi}^2}
   \sum_f  \frac{dN_{\gamma}^f}{dE}(E) B_f\; J(\Theta, \Delta\Omega)\;,
\label{eq:gammaflux}
\end{equation}
where $(\sigma v)$ is the WIMP annihilation rate at zero temperature, $M_{\chi}$ the WIMP
mass and the sum is over all kinematically allowed annihilation final states $f$, each
with a branching ratio $B_f$. It is beyond the scope of the present analysis to review
the ranges of values and the model-dependent determination of these parameters, as well
as of the gamma-ray spectral distributions $dN_{\gamma}^f/dE$, topics which have been
vastly discussed in the recent literature; we will mainly refer here either to a
toy-model in which we pick particular values for $M_{\chi}$ and $(\sigma v)$, and assume
to have only one dominant annihilation channel: it is useful to consider the case of a
soft annihilation channel such as $b-\bar{b}$ pair, and to contrast it with the hard
$\tau^+-\tau^-$ final state. As we showed in \cite{Colafrancesco:2005ji}, these toy
models are well justified in the context of solidly motivated theoretical grounds, for
instance within the paradigm of neutralino dark matter. For definiteness, and for
illustrative purposes, we shall also make use of special, well-studied, benchmark
supersymmetric models, as we did for the case of our analysis of the multi-wavelength
emissions from Coma in Ref.~\cite{Colafrancesco:2005ji}.

\begin{figure*}[!t]
\begin{center}
\includegraphics[scale=0.55]{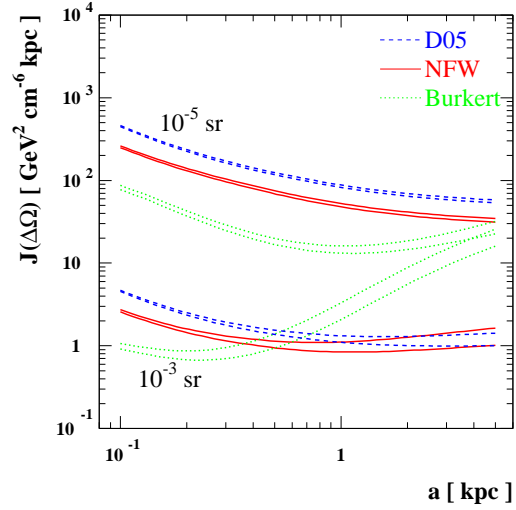}
\end{center}
\caption{We display, for the minimum $\chi^2_{red}$ models illustrated in
Figs.~\protect{\ref{fig:halo1}} and \protect{\ref{fig:halo2}}, the line of sight integral
function $J$,  towards the center of Draco and for two different angular acceptance
$\Delta\Omega$.
} \label{fig:jpsi}
\end{figure*}

In Eq.~(\ref{eq:gammaflux}) the dependence on the halo profile has been factorized out defining:
\begin{equation}
J(\Theta, \Delta\Omega) = \frac{1}{\Delta\Omega} \int_{\Delta\Omega} d\Omega
\int_{l.o.s.} dl \; \rho^2(l)\;,
\end{equation}
where $\Theta$ is the direction of observation and the average is over the angular
acceptance (or the angular resolution) of the detector $\Delta\Omega$. In
Fig.~\ref{fig:jpsi} we plot the range of the expected values for $J$ towards the center
of Draco, for two sample values of $\Delta\Omega$ and within the minimum $\chi^2$ halo
models selected in the previous Section. Confirming other recent
analyses~\cite{Evans:2003sc,Bergstrom:2005qk,Profumo:2006hs}, our results show that there
is a very small spread in the prediction for $J$, even referring to significantly
different dark matter halo shapes and even for small angular acceptances: within a factor
of few and in units of GeV$^2$~cm$^{-6}$~kpc, $J$ is about 100 for $\Delta\Omega =
10^{-5}$~sr and about 1 for $\Delta\Omega = 10^{-3}$~sr. Such small  spread is in
contrast to what one finds in the analogous estimate when considering the Milky Way
galactic center as a source of gamma rays from dark matter annihilation. One can  apply
the same procedure of fitting different halo profiles to the Milky Way dynamical
constraints and then extrapolate their radial scalings down to the innermost parsec or
so; the focus is on the eventual sharp dark matter density enhancement which could be
present in the Galactic center region:  for singular profiles the values of $J$ one
derives may be very large, up to about 10$^4$-10$^5$  for NFW profiles and $\Delta\Omega
= 10^{-5}$~sr (see, e.g.~\cite{Evans:2003sc}; note however that in
Ref.~\cite{Evans:2003sc} a dimensionless $J$ is adopted and to translate values quoted
therein into those for the definition adopted here, one should scale them by the factor
0.765~GeV$^2$~cm$^{-6}$~kpc), but drop dramatically, with a decrease as large as four
orders of magnitude, when considering less singular or cored profiles. in the case of
Draco, the distance of the source is much larger and the l.o.s. integral involves an
average over a large volume, smoothing out the effect of a singularity in the density
profile; at the same time, however, the mean dark matter density is on average fairly
large for any profile, since the dark halo concentration parameter is large.

\begin{figure*}[!t]
\begin{center}
\includegraphics[scale=0.55]{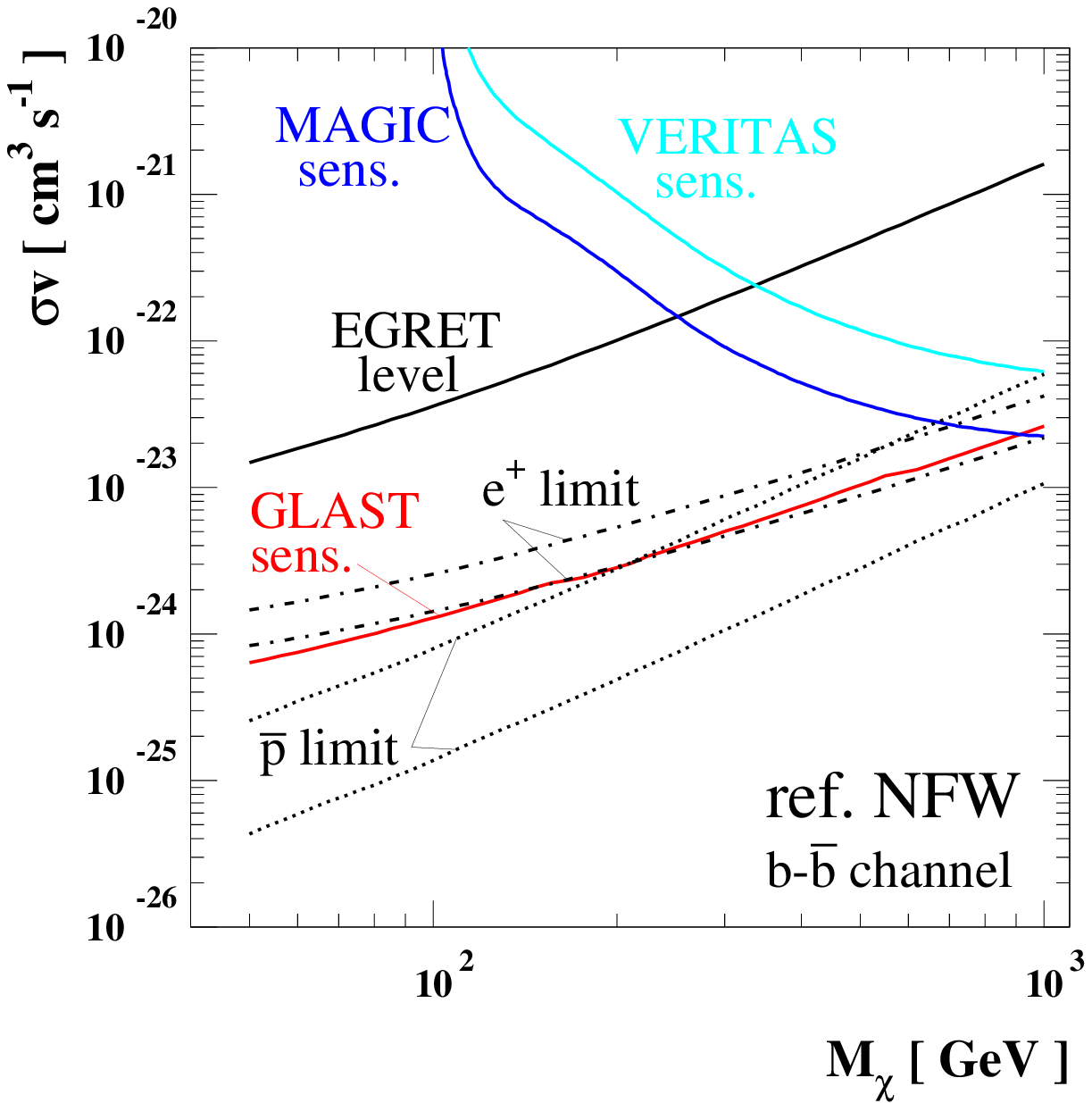}
\quad\includegraphics[scale=0.55]{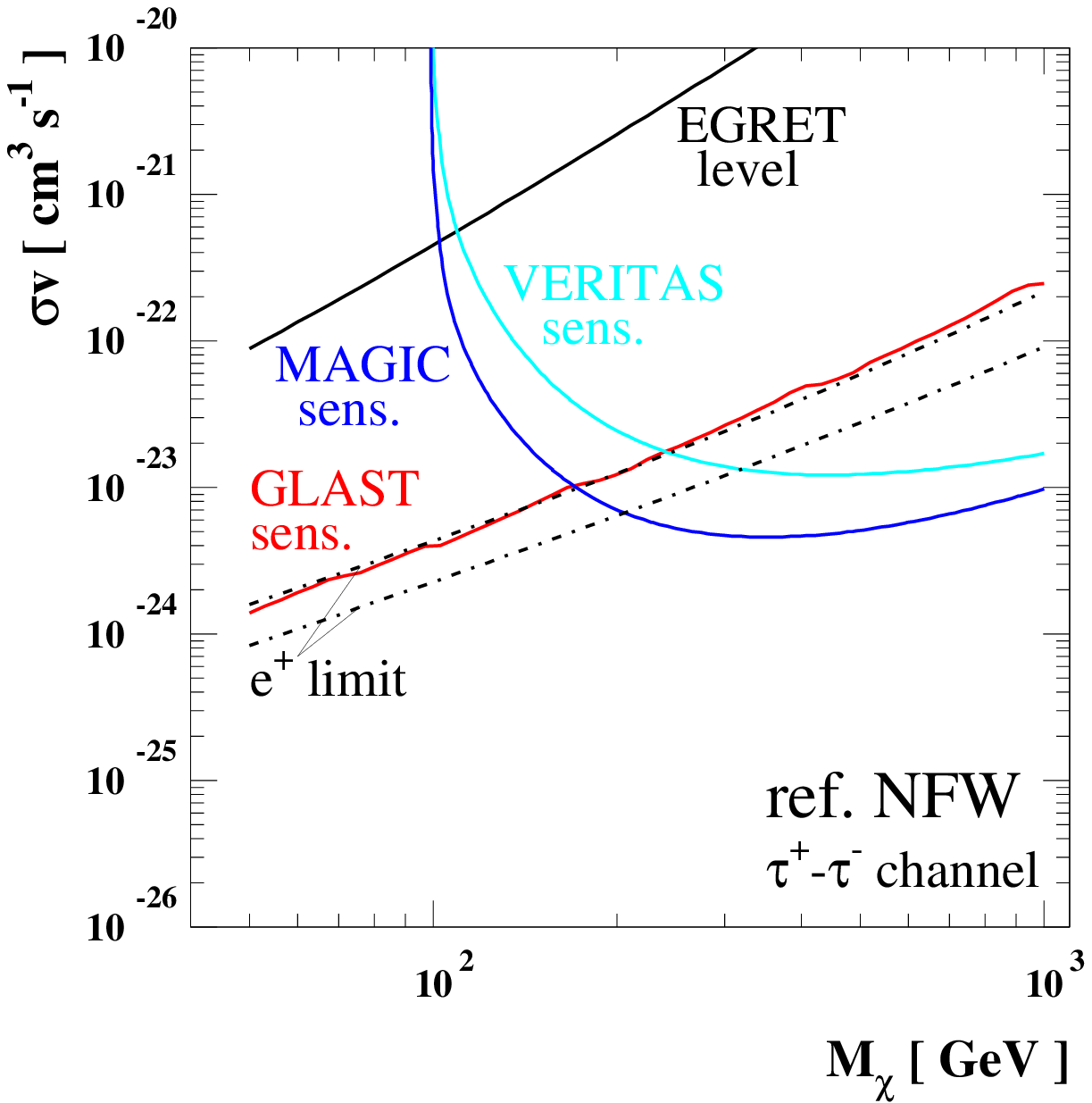}\\
\end{center}
\caption{We show the WIMP mass -- WIMP pair annihilation cross section plane, assuming a $b-\bar{b}$ (left) and $\tau^+-\tau^-$ (right) dominant annihilation final state. The solid black line indicates the value of the annihilation cross section
required for a flux matching the 2 events in EGRET between 1 and 10~GeV, for
exposures and angular cuts as specified in the data analysis,  assuming our reference
NFW best-fit halo model; The black dotted and dot-dashed lines show the limits from the flux of, respectively, antiprotons and positrons in the Milky Way halo (the two lines corresponding to two different cosmic ray propagation setups); The solid red, dark blue and light blue curves indicate the projected sensitivities of GLAST, MAGIC and VERITAS (see the text for more details).}
\label{fig:gamma}
\end{figure*}

In its all-sky survey, EGRET has accumulated a limited exposure towards Draco. A report on
the collected data is given in~\cite{way}; the analysis aims at the identification of a point source
at the center of Draco; seven energy bins are considered, each with the appropriate angular cuts,
no point source is found, and the photon counts are consistent with the expected flux from
diffuse emission, except for a 2 event ``excess'' in the energy range between 1 and 10~GeV,
with a total of 6 events found versus 4.1 expected in standard background scenario
(notice that no statistical evidence for such ``excess'' is claimed in~\cite{way} or in the present analysis).
In Fig.~\ref{fig:gamma} we show, for a given WIMP mass, the value of the annihilation cross section
required for a flux matching the 2 events in EGRET between 1 and 10~GeV, for
exposures and angular cuts as specified in the data analysis,  assuming our reference
NFW best-fit halo model and a $b-\bar{b}$ (left panel) or $\tau^+-\tau^-$  (right panel) annihilation
channels. Also shown in the Figure are expected sensitivity curves with GLAST, the next gamma-ray
telescope in space, and  for upcoming observations of Draco with ground-based ACT telescopes.

Regarding the GLAST detector, we refer to an updated simulation of the instrument
performance~\cite{glastsens}: we refer to the energy dependent sensitivities of the
two LAT sections, the thin (or front) section of the tracker (peak effective area above
1~GeV of about 5500~cm$^2$, 68\% containment angle varying between 0.6 deg at
1~GeV and 0.04 deg at 100~GeV) and of the thick (or back) section of the tracker
(peak effective area above 1~GeV of about 4500~cm$^2$, 68\% containment angle
varying between 1~deg at 1~GeV and 0.07 deg at 100~GeV).  To estimate the background,
we include an extragalactic component  at the level found in EGRET data~\cite{hunter},
extrapolated to higher energy with a $E^{-2.1}$ power law, plus a galactic component scaling
like $E^{-2.7}$ (such scaling is expected  from the decay of $\pi^0$ generated by the
interaction of primary protons with the interstellar medium; we are neglecting an eventual IC
component,  since, if present, such term is most likely already included as misidentified
extragalactic) and normalized in such way that, together with the extragalactic component,
it gives the 6 events above 1~GeV detected by EGRET (assuming just for 4.1 events for the
background level does not change significantly our projected sensitivities).
We consider a 5 years exposure time, in an all-sky survey mode for which the effective area
in the direction of Draco is, on average, about 30\% of the peak effective area (area when
the source is at the zenith of the instrument). Finally, we define a $\chi^2$ variable as:
\begin{equation}
  \chi^2 = \sum_{j =1}^{n_{\rm bins}} \frac{Ns_j^2}{Ns_j+Nb_j}\label{eq:chi2}
\end{equation}
where $Ns_j$ and $Nb_j$ stand for the number of signal and background events in each bin,
restricting to bins with more than 5 signal events. The bin selection should in principle
be optimized model by model; in general we find that it is a good choice to take three
bins per decade in energy (two or more bins are grouped into one in case this procedure
gives 5 signal events; this sometimes happens in the highest energy bin included in the
sum above), while at any given energy we integrate over a solid angle which is the
largest between the PSF set by the 68\% containment angle (full energy dependence
included for each section of the tracker) and the solid angle which maximizes the ratio
$\Phi_s/\sqrt{\Phi_s+\Phi_b}$. Sensitivity curves are given in Fig.~\ref{fig:gamma} as
3~$\sigma$ discovery limits; the latter are found to be, with the present accurate
modeling of the detector, slightly less promising than the analogous curves obtained in
other recent estimates~\cite{Bergstrom:2005qk,Profumo:2006hs}.

Regarding Air Cherenkov Telescopes (ACTs), we consider the detection prospects with
instruments in the northern hemisphere, i.e. MAGIC \cite{magic}, which is currently
taking data, and VERITAS, which will be completed soon extending the current single
mirror telescope to an array of at first four, then later seven telescopes
\cite{veritas}. We assess the discovery sensitivity of the two ACTs using a low energy
threshold of 100 GeV, and the effective collection area as a function of energy recently
quoted by the two collaboration in Ref.~\cite{magic,veritas}; the main sources of
background for ACTs correspond to misidentified gamma-like hadronic showers and
cosmic-ray electrons. The diffuse gamma-ray background gives a subdominant contribution
to the background, which we also took into account using the same figures outlined above
for the space-based telescopes background. We use the following estimates for the ACT
cosmic ray background \cite{pierobuck}:
\begin{eqnarray}
     \frac{{\rm d}N_{\rm had}}{{\rm d}\Omega}(E>E_0)&=&6.1\times
     10^{-3}\epsilon_{\rm had}\left(\frac{E_0}{1\ {\rm
     GeV}}\right)^{-1.7}\srunits,\label{eq:had}\\
     \frac{{\rm d}N_{\rm el}}{{\rm d}\Omega}(E>E_0)&=&3.0\times
     10^{-2}\left(\frac{E_0}{1\ {\rm
     GeV}}\right)^{-2.3}\srunits,\label{eq:el}
\end{eqnarray}
where $\epsilon_{\rm had}$ parameterizes the efficiency
of hadronic rejection, which we assume to be at the level of 10\% \cite{magic,veritas}. As above, we proceed with an optimized binning of the energy range of interest (extending from the energy threshold up to the WIMP mass), compute the number of signal and background events in each bin, and require that the resulting $\chi^2$ (evaluated according to the analogue of Eq.~(\ref{eq:chi2})) gives a statistical excess over background.

The models for which we predict a detectable flux have fairly large cross sections, still
compatible but in the high end of models with a thermal relic density, as computed in a
standard cosmological scenario, which matches the observed dark matter density in the
Universe~\cite{WMAP}, see, e.g., \cite{Colafrancesco:2005ji} (another possibility is that
we refer to models with non-thermal relic components, such as from the decay of moduli
fields, or to modified cosmological setups affecting the Hubble parameter at the time of
WIMP decoupling, see, e.g., ~\cite{non-thermal}). Large annihilation cross sections give
enhanced signals for any indirect dark detection technique, in particular we need to
check whether this picture is compatible with the antimatter fluxes measured at Earth: in
fact, pair annihilation of WIMP in the halo of the Milky Way is acting as  a source of
primary positrons and antiprotons which diffuse in the magnetic field of the Galaxy,
building up into exotic antimatter populations. Current measurements of the local
antiproton flux are consistent with the standard picture of antiprotons being secondary
particles generated by the primary cosmic ray protons in spallation
processes~\cite{bess}; on the other hand, a weak evidence of an excess in the positron
flux has been claimed~\cite{heat,ams}, in a picture that is going to become increasingly
clearer with the on-going measurements in space by the recently launched Pamela
detector~\cite{pamela}. We estimate the induced flux of positrons and antiprotons (no
antiprotons are generated in the $\tau^+-\tau^-$ channel), referring to the  same Milky
Way halo model we have introduced to estimate tidal effects on Draco, and to the
diffusive convective model for the propagation of charged particles implemented in the
DarkSUSY package~\cite{ds}. Parameters in the propagation model are chosen in analogy
with a standard setup \cite{Profumo:2004ty} in the GALPROP propagation
package~\cite{galprop}, or the most conservative choice suggested in Ref.~\cite{salati}
which can still reproduce ratios of secondaries to primaries as measured in cosmic ray
data while minimizing the flux induced by WIMP annihilations:
these give, respectively, the lower and upper curves plotted in Fig.~\ref{fig:gamma} and
corresponding to the 3~$\sigma$ limits on the annihilation cross section obtained by
comparing the WIMP-induced fluxes to a full compilation of present data on the local
antiproton and positron fluxes. The values displayed should should not be taken as strict
exclusion curves, since we are not doing a full modeling of the uncertainties in the
propagation model, nor scanning on more general configurations of the Milky Way dark
matter halo; relaxing them by a factor of 2 to 5 or maybe even larger should be
relatively straightforward; they can be however taken as a guideline to see that models
in such region of the parameter space should be testable with higher precision antimatter
data, while models at the EGRET level are most probably already excluded by current data.

\begin{figure*}[!t]
\begin{center}
\includegraphics[scale=0.55]{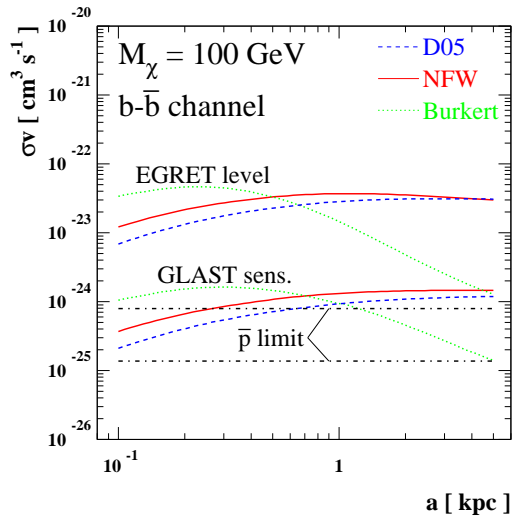}\\
\end{center}
\caption{As in Fig.~\protect{\ref{fig:gamma}}, but keeping the WIMP mass fixes at 100
GeV, and varying the halo scale length parameter $a$ for the three dark matter density
profiles considered here.
}
 \label{fig:gamma2}
\end{figure*}

Finally in Fig.~\ref{fig:gamma2} we show how flux levels and projected sensitivities scale
scanning over our sample of halo models for Draco, for a sample WIMP mass and the
$b-\bar{b}$ final state; antiproton limit levels are also plotted for comparison.

This work which is based on realistic estimates of both the dark matter density
distribution in Draco and of the WIMP physical set up leaves the early Cactus claims
(see, e.g., http://ucdcms.ucdavis.edu/solar2/results/Chertok.PANIC05.pdf) of a detection
of gamma-ray emission from Draco apart.

\section{Multi-wavelength signals from Draco}

The following step is to extend our analysis to the radiation emitted at lower frequencies.
For this purpose, we need to track the injection of electrons and positrons from WIMP
annihilations in Draco, as well as their propagation in space and energy; it will then be possible
to make predictions for the induced synchrotron  and Inverse Compton radiations.
Our starting point is the assumption that, in analogy to more massive objects such as the
Milky Way itself, there is a random component of interstellar magnetic fields  associated with
Draco and that it is a fair approximation to model the propagation of charged particles as
a diffusive process. In this limit we can calculate the electron and positron number densities
implementing  to the following transport equation:
\begin{equation}
  \frac{\partial}{\partial t}\frac{dn_e}{dE} =
  \nabla \left[ D(E,\vec{x}) \nabla\frac{dn_e}{dE}\right] +
  \frac{\partial}{\partial E} \left[ b(E,\vec{x}) \frac{dn_e}{dE}\right]+
  Q_e(E,\vec{x}) \;,
\label{diffeq}
\end{equation}
where $Q_e$ is the electron or positron source function from WIMP annihilations:
\begin{equation}
  Q_e(E,\vec{x}) =  \frac{1}{2\,M_{\chi}^2}
   \sum_f  \frac{dN_{e}^f}{dE}(E) B_f\; \rho^2(\vec{x})\;,
\label{eq:epmsource}
\end{equation}
while $D$ is the diffusion coefficient and $b$ the energy loss term.
In Ref.~\cite{Colafrancesco:2005ji} we derived the analytic
solution to this equation in case of a spherical symmetric system and for $D$ and $b$ that
do not depend on the spatial coordinates. We refer here to a time-independent source and
consider the limit for an electron number density that has already reached equilibrium;
the solution takes the form:
\begin{equation}
\frac{dn_e}{dE}\left(r,E \right) =  \frac{1}{b(E)} \int_E^{M_\chi}  dE' \;
\widehat{G}\left(r,v-v' \right) Q_e(r,E')
\label{eq:full2}
\end{equation}
with:
\begin{eqnarray}
\widehat{G}\left(r, \Delta v\right) & = &
\frac{1}{[4\pi(\Delta v)]^{1/2}}
\sum_{n=-\infty}^{+\infty} (-1)^n
\int_0^{r_h} dr' \frac{r'}{r_n} 
\left[\exp{\left(-\frac{(r'-r_n)^2}{4\,\Delta v}\right)}-
\exp{\left(-\frac{(r'+r_n)^2}{4\,\Delta v}\right)}\right]
\frac{\rho^2(r')}{\rho^2(r)}\;.
\label{eq:rescaling}
\end{eqnarray}
In the Green function $\widehat{G}$ the energy dependence has been hidden in two subsequent
changes of variable $v=\int_{u_{\rm min}}^u d\tilde{u} D(\tilde{u})$ and
$u =\int_E^{E_{\rm max}} \frac{dE^{\prime}}{b(E^{\prime})}$; the radial integral extends up to the
radius of the diffusion zone $r_h$ at which a free escape boundary condition is imposed, as enforced
by the sum over $n$, having defined $r_n = (-1)^n r + 2 n r_h$. Whenever the scale of mean
diffusion $\sqrt{\Delta v}$, covered by an electron while losing energy from energy at the source
$E^\prime$ to the energy when interacting $E$, is much smaller than the scale over which $\rho^2$
has a significant variation, $\widehat{G}$ is close to 1 and spatial diffusion can be neglected;
in Ref.~\cite{Colafrancesco:2005ji} it was shown that this limit applies in the case of the Coma cluster. For dSph we find that, most likely, we are in the opposite regime.

\begin{figure*}[!t]
\begin{center}
\includegraphics[scale=0.55]{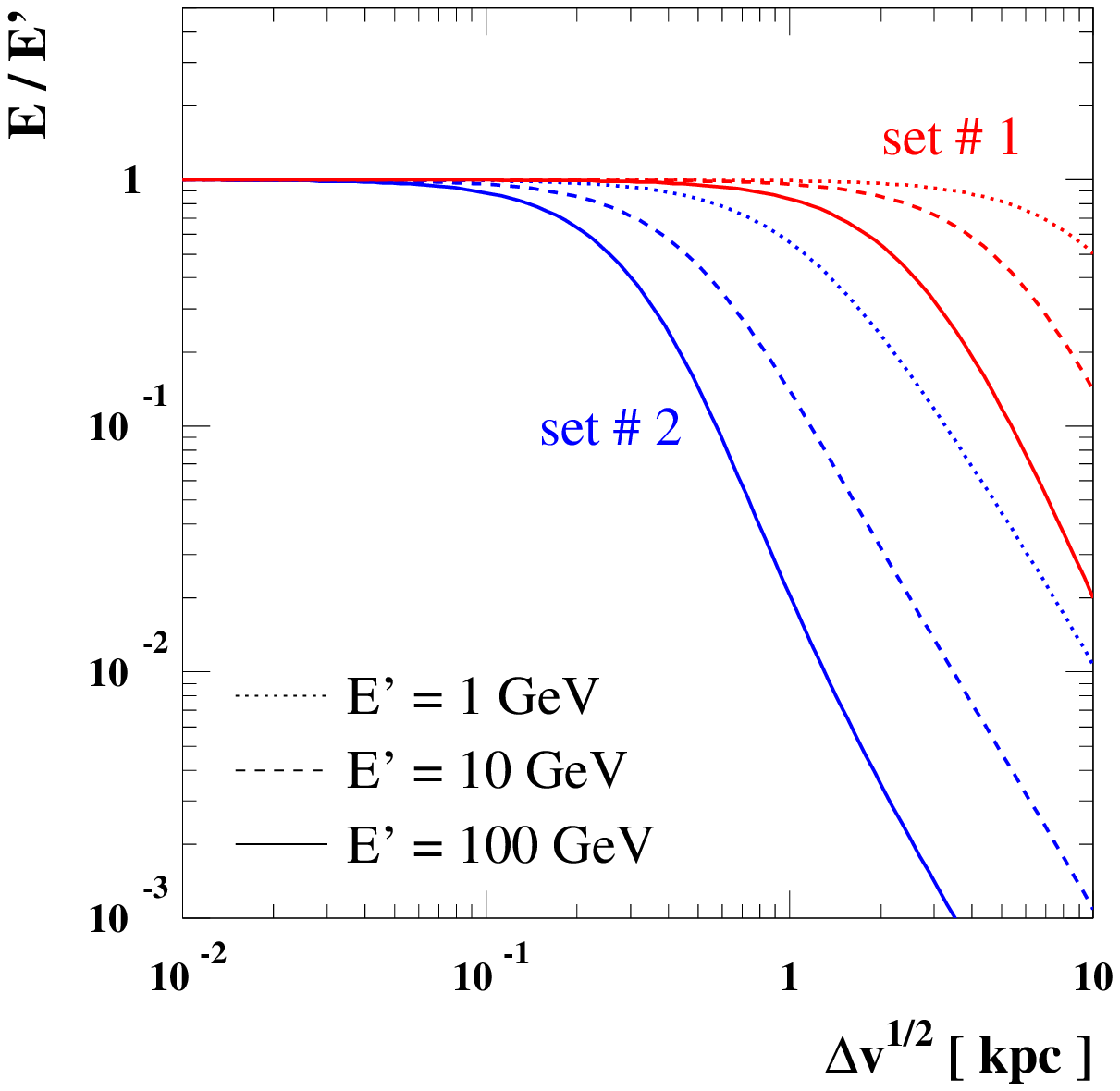}
\quad\includegraphics[scale=0.55]{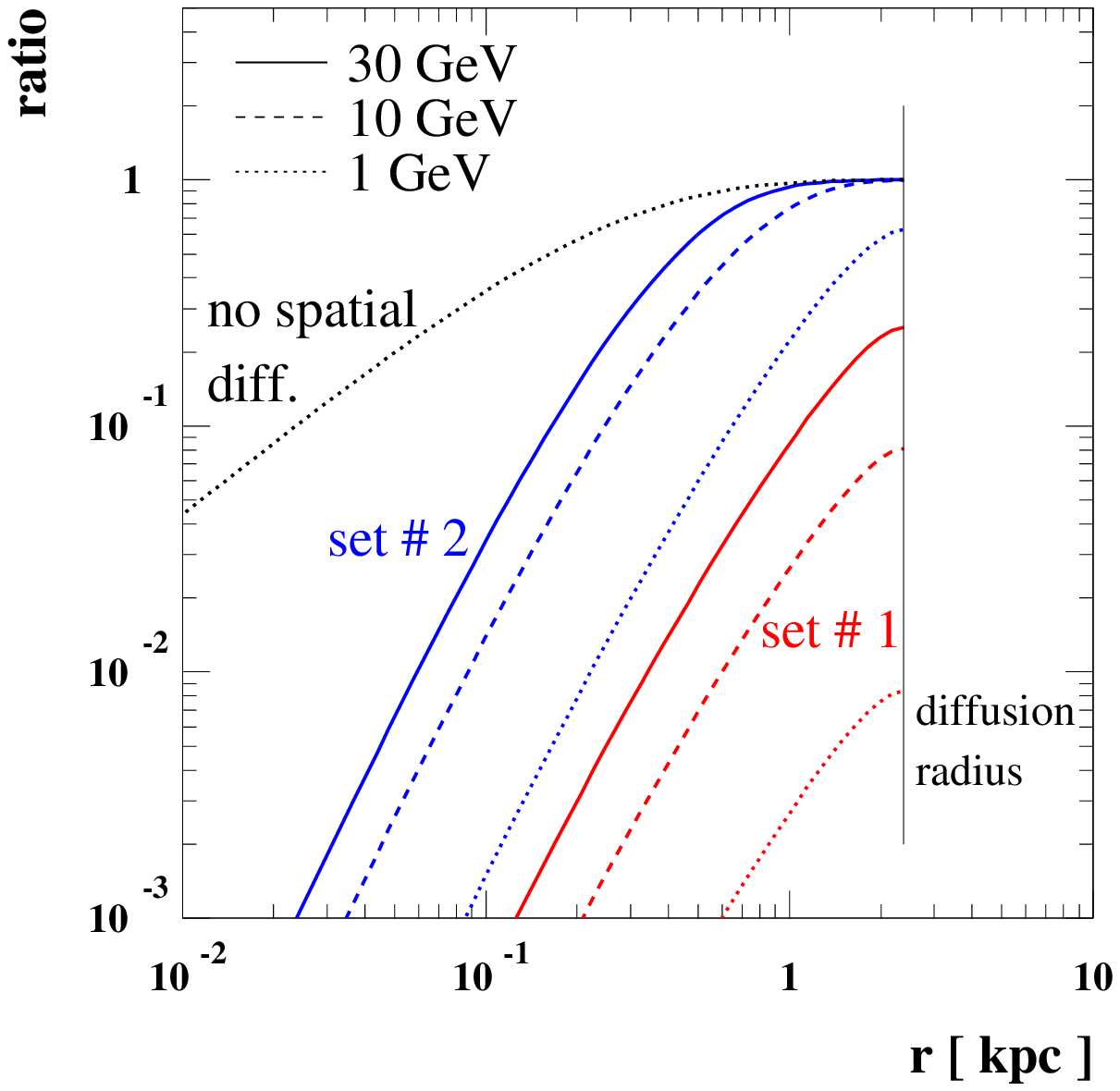}\\
\end{center}
\caption{Left panel: mean diffusion distance $\sqrt{\Delta v}$ covered by an electron while
losing energy from its energy at emission $E^\prime$ to its energy at interaction $E$; three
different $E^\prime$ are considered, as well as a conservative choice for propagation
parameters (labeled by "set \# 1" in the plot) or a more extreme choice (labeled by "set \# 2" in
the plot),  see the text for details. Right panel: we plot, as function of the distance from the center
of Draco $r$, the integral over volume up the radius $r$ of the electron number density
$dn_e/dE(r,E)$, for three values of the energy $E$ and for the two choices of propagation
parameters already considered in the left panel; also shown is the case when $dn_e/dE$
is computed assuming negligible spatial diffusion. For each energy, integrals over volume
are normalized to the integral over volume up to the assumed radius of diffusion for Draco
of  $dn_e/dE$ in the limit in which spatial diffusion is neglected; we have chosen the
reference NFW model for the dark halo, and a WIMP of mass 100~GeV annihilating into
$b\,\bar{b}$. As it can be seen, compared to the case when spatial diffusion is neglected,
in the actual cases applying to Draco there is a sharp deficit of electrons in the inner region
of even within the total diffusion volume.
} \label{fig:diffusion}
\end{figure*}

To model electron and positron energy losses we choose as a reference value for the
magnetic field $B_{\mu} = 1$~$\mu$G and a thermal electron density of
$10^{-6}$~cm$^{-3}$: these values are derived from radio observations of dwarf galaxies
similar to Draco at $5 $ GHz (Klein et al. 1992) and from the assumption that the ROSAT
PSPC X-ray upper limit on Draco (Zang \& Meurs 2001) is due to thermal bremsstrahlung,
respectively.
For the diffusion coefficient we assume the Kolmogorov form $D(E) =  D_0/B_\mu^{1/3}
\left(E/{\rm{1\;GeV}}\right)^{1/3}$, fixing the constant $D_0 = 3 \cdot
10^{28}$~cm$^{2}$~s$^{-1}$ in analogy with its value for the Milky Way; finally our guess
for the dimension of the diffusion zone is that, again consistently with the picture
relative to the Milky Way,  $r_h$ is about twice the radial size of the luminous
component, i.e., here, 102~arcmin (we will refer to this set of propagation parameters as
set \#1). In the left panel of Fig.~\ref{fig:diffusion} we show that, in this setup,
electrons and positrons lose a moderate fraction of their energies on scales
$\sqrt{\Delta v}$  that are comparable to the size of the diffusion region, i.e. spatial
diffusion as a large effect. Even  referring to an extreme model in which the diffusion
coefficient is decreased of two orders of magnitudes down to $D_0 = 3 \cdot
10^{26}$~cm$^{2}$~s$^{-1}$ (this would imply a much smaller scale of uniformity for the
magnetic field), adding on top of that a steeper scaling in energy, $D(E) = D_0
\left(E/{\rm{1\;GeV}}\right)^{-0.6}$ (this is the form sometime assumed for the Milky
Way; we label this propagation parameter configuration set \#2),  scales $\sqrt{\Delta
v}$ are decreased but remain still relatively large. In the right panel of
Fig.~\ref{fig:diffusion} we consider a WIMP model with mass 100~GeV annihilating in the
$b\,\bar{b}$ final state within our reference NFW halo model for Draco. We show integrals
over volume within the radial coordinate $r$ of the equilibrium number density $dn_e/dE$,
for a few values of the energy $E$, and for the set of propagation parameters \#1 and
\#2, as well as the results corresponding to the assumption that spatial diffusion can be
neglected. All integrals are normalized to the integrals over the whole diffusion region
of $dn_e/dE$ for the corresponding energy $E$ and assuming negligible spatial diffusion:
we deduce from the figure that for set \#1 there is a depletion of the electron/positron
populations with a significant fraction leaving the diffusion region, while for set \#2
they are more efficiently confined within the diffusion region but still significantly
misplaced with respect to the emission region.

\begin{figure*}[!t]
\begin{center}
\includegraphics[scale=0.55]{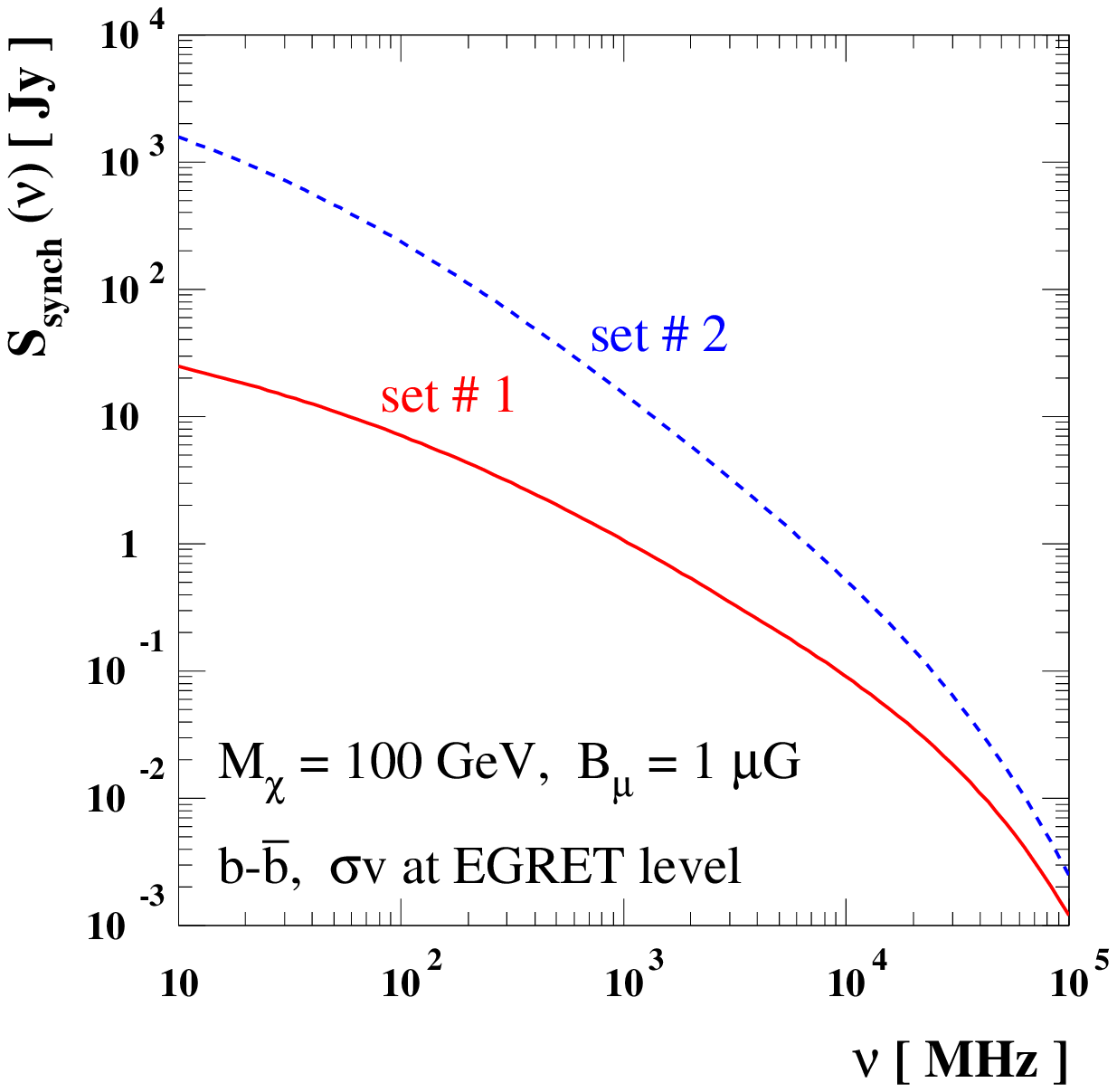}
\quad\includegraphics[scale=0.55]{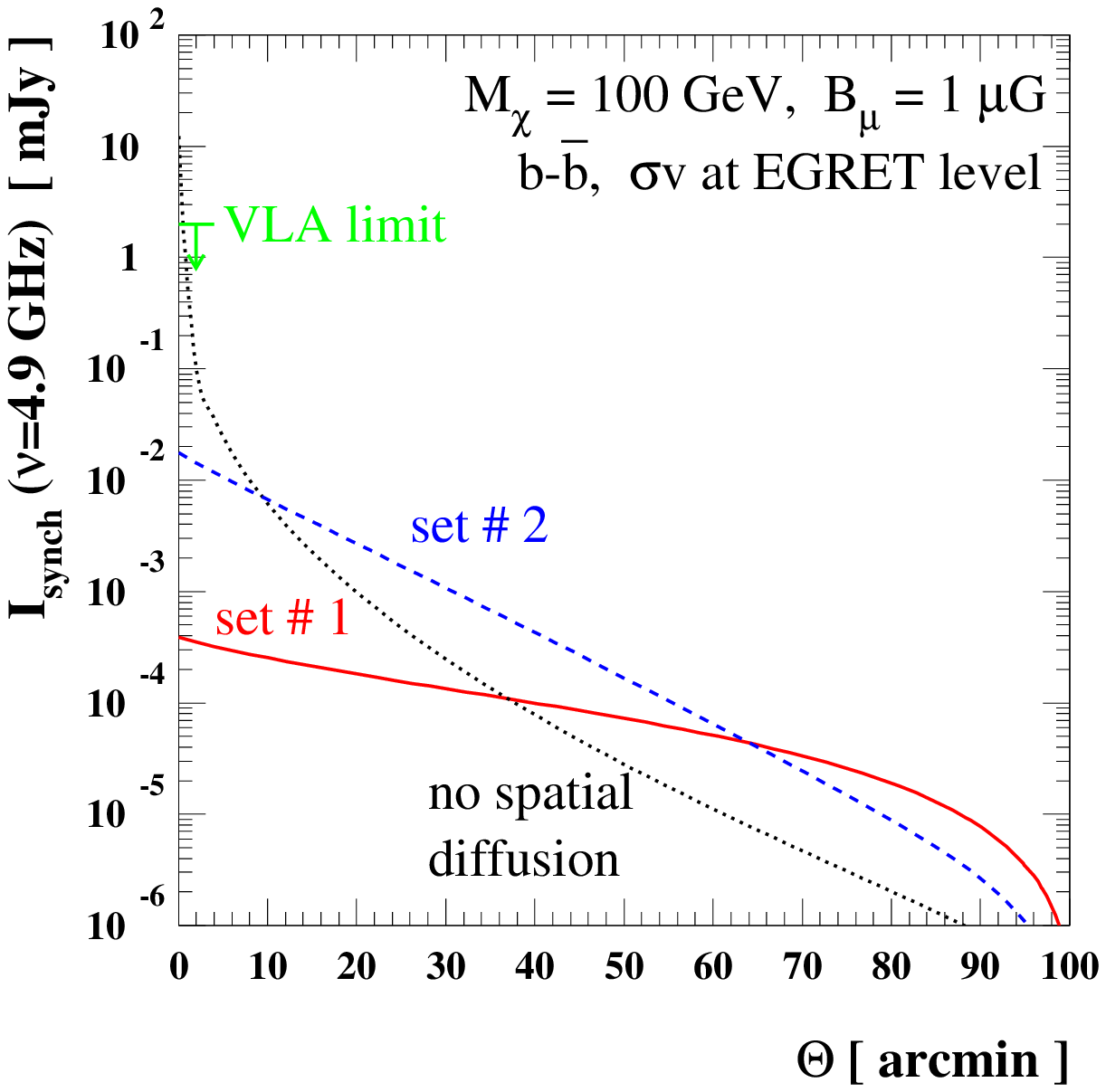}\\
\end{center}
\caption{Radio flux density spectrum (left panel) and surface brightness distribution
at the frequency $\nu = 4.9$~GHz for a sample WIMP model of mass 100~GeV
annihilating into $b\,\bar{b}$ with a cross section tuned at the level to give 2 events
in the EGRET gamma-ray telescope. Results are given for the two choices of propagation
parameters already considered in Ref.~\protect{\ref{fig:diffusion}}, and for the surface
brightness, for illustrative purposes only, in case when spatial diffusion is neglected.
Surface brightness are plotted for  a 3~arcsec angular acceptance, corresponding to the
VLA angular resolution at the time a search with this instrument for a point source
at the center of Draco was performed (the obtained upper limit is plotted in the figure).
} \label{fig:radio}
\end{figure*}

For a given electron/positron equilibrium distribution we can infer the induced synchrotron
and inverse Compton emissions.
In the limit of frequency $\nu$ of the emitted photons much larger than the non-relativistic
gyro-frequency $\nu_0= e B/(2\pi m c) \simeq 2.8 B_{\mu}$~Hz, the spontaneously emitted
synchrotron power takes in the form~\cite{book}:
\begin{equation}
P_{\rm synch}\left(\nu,E,r\right) = \int_0^\pi d\theta \frac{\sin\theta}{2} 2 \pi \sqrt{3} r_0 m c
\nu_0 \sin\theta F\left(x/\sin\theta\right) \;,
\end{equation}
where we have introduced the classical electron radius $r_0= e^2/(m c^2) = 2.82 \cdot 10^{-13}$~cm,
and we have defined the quantities $x$ and $F$ as:
\begin{equation}
x \equiv \frac{2\nu}{3\nu_0 \gamma^2}
\left[ 1+ \left( \frac{\gamma \nu_p}{\nu}\right)^2\right]^{3/2}\;,
\end{equation}
and
\begin{equation}
F\left(t\right) \equiv t \int_t^\infty dz K_{5/3}(z) \simeq
1.25 t^{1/3} \exp{(-t)} \left[ 648 + t^2\right]^{1/12}\;.
\end{equation}
Folding the synchrotron power with the spectral distribution of the equilibrium number
density of electrons and positrons, we find the local emissivity at the frequency $\nu$:
\begin{equation}
j_{\rm synch}\left(\nu,r\right) = \int_{m_e}^{M_{\chi}} dE\, \left(\frac{dn_{e^-}}{dE} +
\frac{dn_{e^+}}{dE} \right) P_{\rm synch}\left(\nu,E,r\right)\;.
\end{equation}

In Fig.~\ref{fig:radio} we consider a reference WIMP model with a mass of 100~GeV,
annihilating into $b-\bar{b}$ with a cross section at the level to induce a gamma-ray flux
matching the 2 events in EGRET between 1 and 10~GeV for a dark matter distribution
as in our reference NFW model. In the left panel we  plot the radio flux density spectrum
integrated over the whole diffusion volume:
\begin{equation}
S_{\rm synch}(\nu)= \int d^3r \, \frac{ j_{\rm synch}\left(\nu,r\right)}{4 \pi\, d^2}\;,
\end{equation}
with $d$ the distance of Draco. The spectrum is significantly flatter for the propagation
parameter set \#1, since the peak in emitted photon frequency is linearly proportional to the
energy of the emitting particle, and electrons and positrons tend to escape from the diffusion
box while losing energy, rather than remaining confined within it and giving a signal which
piles up at lower frequencies. We then introduce the azimuthally averaged surface
brightness distribution:
\begin{equation}
I_{\rm synch}(\nu,\Theta,\Delta\Omega)= \int_{\Delta\Omega} d\Omega \int_{l.o.s.} dl \,  \frac{j_{\rm
synch}\left(\nu,l\right)}{4 \pi} \;,
\end{equation}
where the integral is performed along the l.o.s. $l$, within a cone of angular
acceptance $\Delta\Omega$ centered in a direction forming an angle $\Theta$ with the direction
of the center of Draco. In the right panel of Fig.~\ref{fig:radio} we plot surface brightness
integrated over a cone of 3~arcsec width, corresponding to the tiny angular acceptance
of the VLA at the time it was used to perform a searches for point radio sources in the central
4~arcmin of Draco~\cite{Fomalontetal1979}; no source was found and the corresponding upper limit
is plotted in the figure. To illustrate how the shape of the signal changes compared to that
of the source function, we also plot the surface brightness which  we would obtain in the limit
of no spatial diffusion. In Ref.~\cite{tyler} radio fluxes are computed in this limit and the VLA
measurement is exploited to exclude WIMP models; we have demonstrated in our discussion
that the limit of no spatial diffusion is not likely to hold in case of Draco and the figure
illustrates the fact that, with present data, limits on the model stemming from radio
frequencies are less constraining than in the gamma-ray band. The figure illustrates also
another point: the gamma-ray flux retraces the WIMP annihilation source function; in the example
we have considered, even with the angular resolution at which future observations will be carried
out, Draco would appear as a single point source. On the other hand, in the radio band the
signal is spread out  over a large angular size, standing clearly as diffuse emission.

\begin{figure*}[!t]
\begin{center}
\includegraphics[scale=0.55]{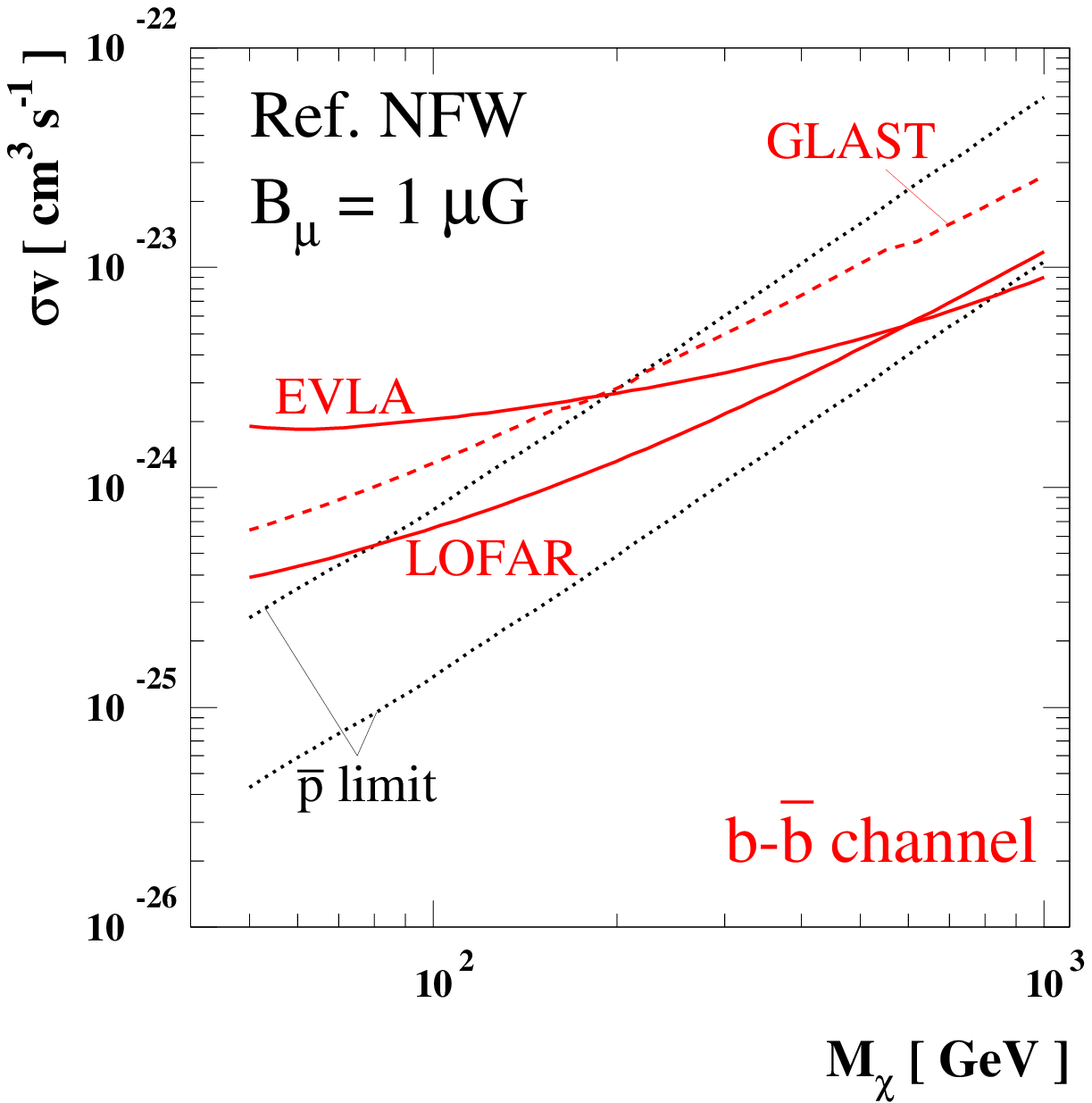}
\quad\includegraphics[scale=0.55]{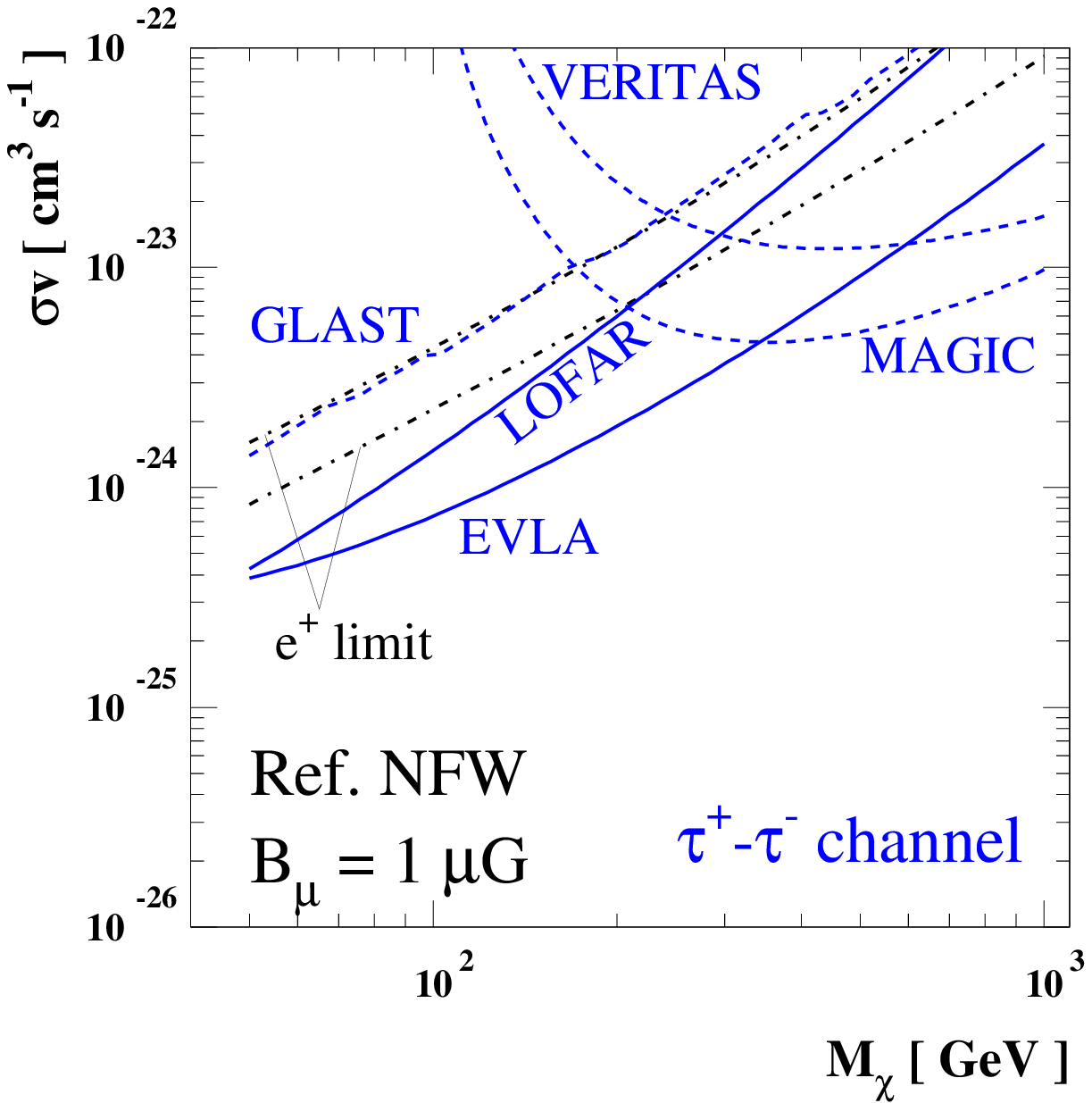}\\
\end{center}
\caption{The projected sensitivity of future diffuse radio source searches from the direction of Draco, in the same setup as in Fig.~\protect{\ref{fig:gamma}}. For comparison, we also indicate the sensitivity of gamma-ray search experiments, and the constraints from the antiproton and positron fluxes (the two lines corresponding to two different propagation setups for the Milky Way).}
\label{fig:radiolim}
\end{figure*}

\begin{figure*}[!t]
\begin{center}
\includegraphics[scale=0.55]{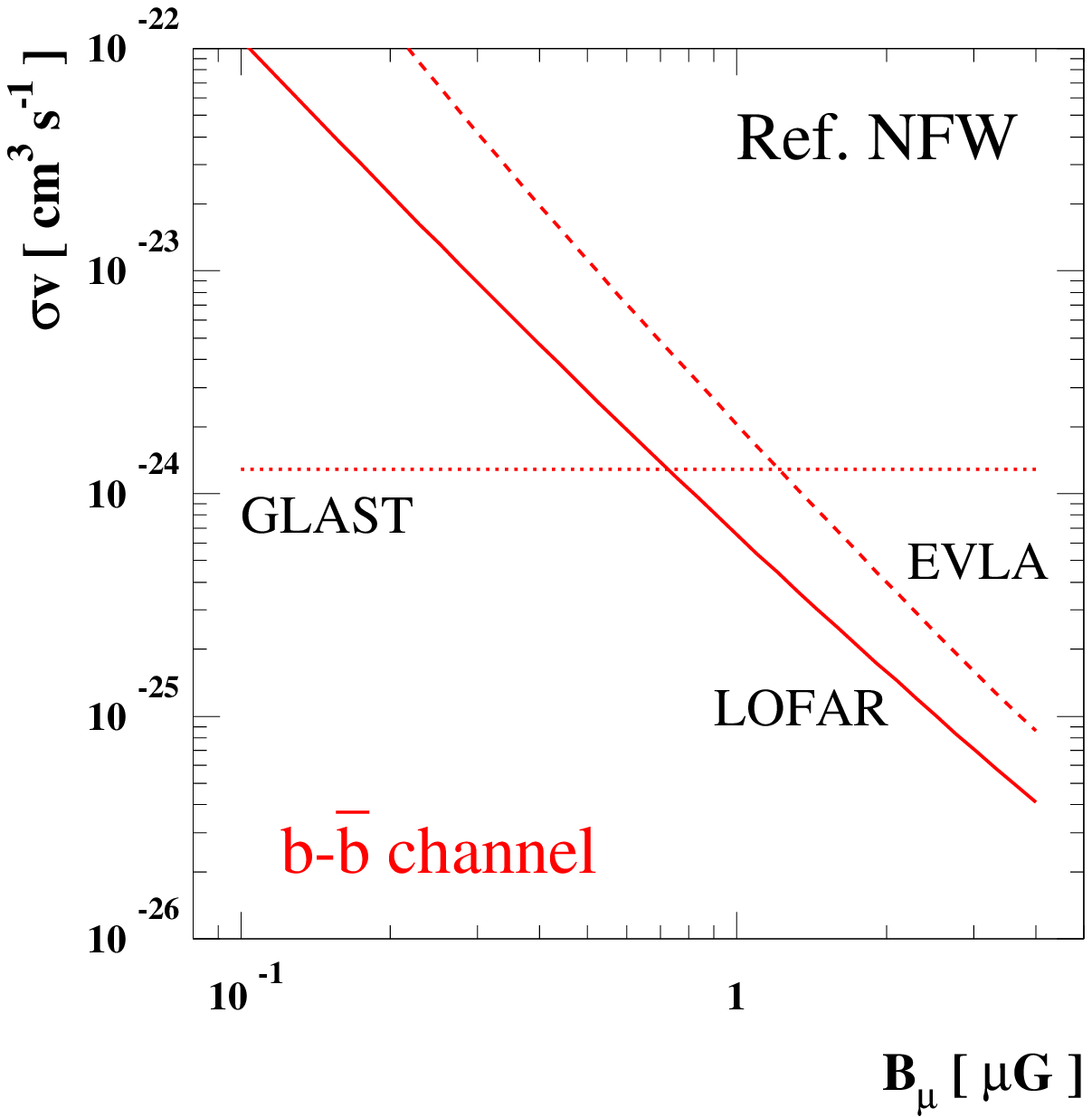}
\quad\includegraphics[scale=0.55]{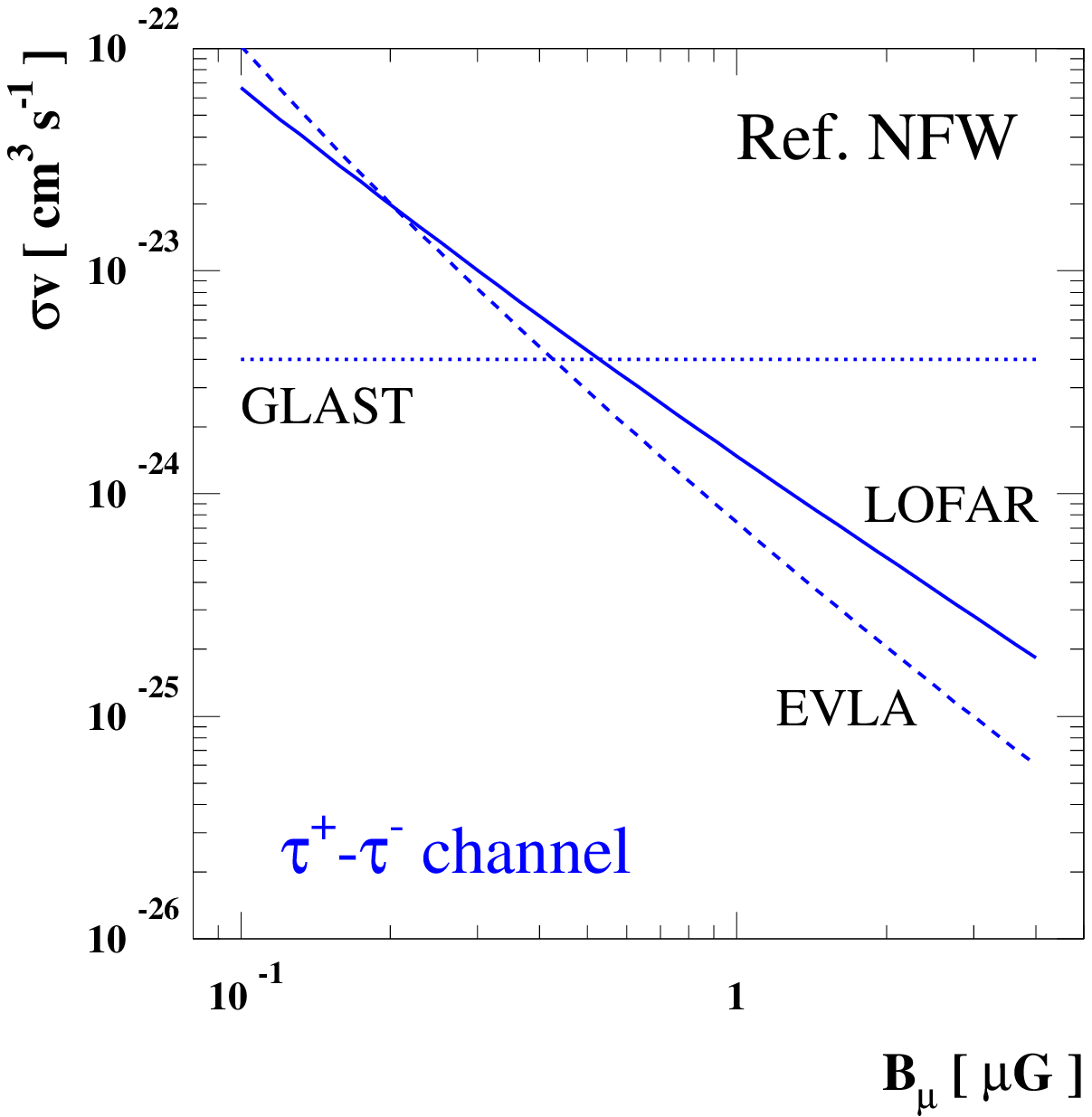}\\
\end{center}
\caption{The projected sensitivities of future searches for a diffuse radio source as a function of the magnetic field in Draco, in the same setting as in  Fig.~\protect{\ref{fig:gamma}}, with a WIMP mass set to 100 GeV. For comparison, we also show the projected sensitivity of GLAST.}
\label{fig:radiolimb}
\end{figure*}

\begin{figure*}[!t]
\begin{center}
\includegraphics[scale=0.55]{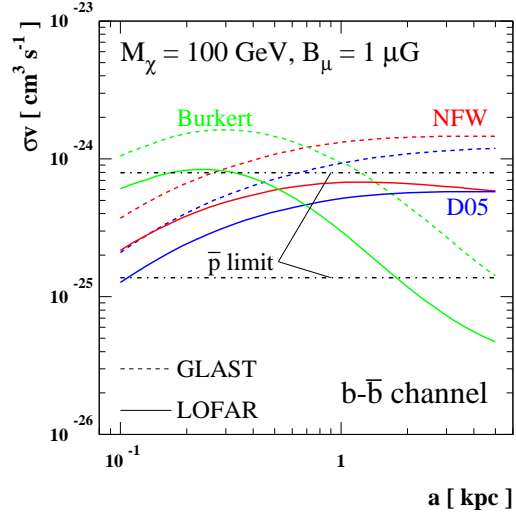}
\end{center}
\caption{LOFAR projected sensitivities, for a given WIMP model and value of the magnetic field
in Draco, as a function of the length scale parameter $a$ in the class of halo models selected
in Sec.~\protect{\ref{sec:halo}}, compared to the GLAST projected sensitivity.}
\label{fig:radiolimc}
\end{figure*}

No search for a diffuse radio emission from Draco has been performed so far.  Even with a future next generation radio telescope the quoted sensitivities do not apply for an extended source. To
address the discovery potential of such apparata for the signal we are analyzing, we make a simple extrapolation on the quoted point source sensitivity
$I_p(\nu)$ for a reference angular resolution $\Delta\Omega_p$ and integration time
$\Delta t_p$, assuming a homogeneous background:
\begin{equation}
I_{min}(\nu,\Delta\Omega) = I_p(\nu) \frac{\sqrt{\Delta\Omega}}{\sqrt{\Delta\Omega_p}}
 \frac{\sqrt{\Delta t_p}}{\sqrt{\Delta t}}\;.
\end{equation}
Reference values for  EVLA in phase I~\cite{evla}, at $\nu = 5$~GHZ are
$I_p(\nu) = 0.8 \; \mu$Jy for $\Delta t_p = 12$~hr and $\Delta\Omega_p = 0.4$~arcsec;
for LOFAR observations are al lower frequency up to $\nu = 200$~MHZ, for which
$I_p(\nu) = 0.03$~mJy with $\Delta t_p = 1$~hr and $\Delta\Omega_p = 0.64$~arcsec \cite{lofar}.
To infer the projected sensitivity limit, for each WIMP model, halo profile and frequency
of observation, we compute the value for the angular acceptance $\Delta\Omega$ at which
$I_{\rm synch}(\nu,\Delta\Omega)/\sqrt{\Delta\Omega}$ is maximal; we also assume as
exposure time $\Delta t = 8$~hr.  In Fig.~\ref{fig:radiolim} we show results for the sensitivity
curves in the plane annihilation cross section versus WIMP mass, for our reference
NFW profile, the conservative set \#1 for propagation parameters (set \#2 would give
much more favorable results)
and a value of the magnetic field equal to 1~$\mu$G.  The figure shows that, for this
choice of parameters, we predict that WIMP models that are at the level of being detected
by GLAST  or ACTs in gamma-rays, should also give a detectable radio flux, possibly with
an even higher sensitivity in favorable propagation configuration. However, this last
conclusion is more model dependent, since some of the parameters are crucial for the
estimate of the radio flux. In Fig.~\ref{fig:radiolimb} we illustrate this point, fixing
the WIMP mass to 100~GeV, varying instead the value of magnetic field in Draco: the trend
is obviously that the larger the magnetic field, the higher the induced radio flux, but
the dependence is not trivial since the magnetic field enters both in the propagation of
electrons and positrons, and in the emission of synchrotron radiation. Finally in
Fig.~\ref{fig:radiolimc} we examine the dependence of the radio sensitivity curves on the
model describing the dark matter halo in Draco: we find scalings that are analogous,
although different in fine details,  to those sketched previously for the gamma-ray
fluxes  and the corresponding sensitivity of the GLAST satellite.

The inverse Compton emission on the cosmic microwave background and on starlight
fills the gap in frequency between radio and gamma-ray frequencies. Let $E= \gamma m_e
c^2$ be the energy of electrons and positrons, $\epsilon$ the target photon energy and
$E_\gamma$ the energy of the scattered photon; the Inverse Compton power is obtained
by folding the differential number density of target photons with the IC scattering cross section:
\begin{equation}
P_{\rm IC}\left(E_\gamma,E\right) = c \, E_\gamma \int d\epsilon \, n(\epsilon) \,
\sigma(E_\gamma, \epsilon, E)
\label{eq:ICpower}
\end{equation}
where $\sigma(E_\gamma, \epsilon, E)$ is the Klein-Nishina cross section~\cite{book} and
$n(\epsilon)$ is the differential energy spectrum of the target photons; for simplicity
we will assume that the starlight spectrum has the shape of a black body with temperature
$T= 0.3$~eV
Such value of the temperature has been estimated on the basis of the fact that the major
part of the Draco star are halo stars somewhat below the turnoff point of the subdwarf
main sequence (Odenkirchen et al. 2001). The effective temperature in the HR diagram with
Fe/H=-2.0 dex and with an age of $\sim 12$ Gyr is of the order of $T \approx 3300-3500$
K, which is equivalent to an energy of $\approx 0.28-0.3$ eV.
Folding the IC power with the spectral distribution of the equilibrium number
density of electrons and positrons, we get the local emissivity of IC photons of energy
$E_\gamma$:
\begin{equation}
j_{\rm IC}\left(E_\gamma, r\right) = \int
dE\, \left(\frac{dn_{e^-}}{dE} + \frac{dn_{e^+}}{dE} \right)
P_{\rm IC}\left(E_\gamma,E\right)\;
\label{eq:ICemiss}
\end{equation}
and the azimuthally averaged surface brightness distribution:
\begin{equation}
I_{\rm IC}(\nu,\Theta,\Delta\Omega)= \int_{\Delta\Omega} d\Omega \int_{l.o.s.} dl \,  \frac{j_{\rm
IC}\left(\nu,l\right)}{4 \pi} \;.
\end{equation}

\begin{figure*}[!t]
\begin{center}
\includegraphics[scale=0.55]{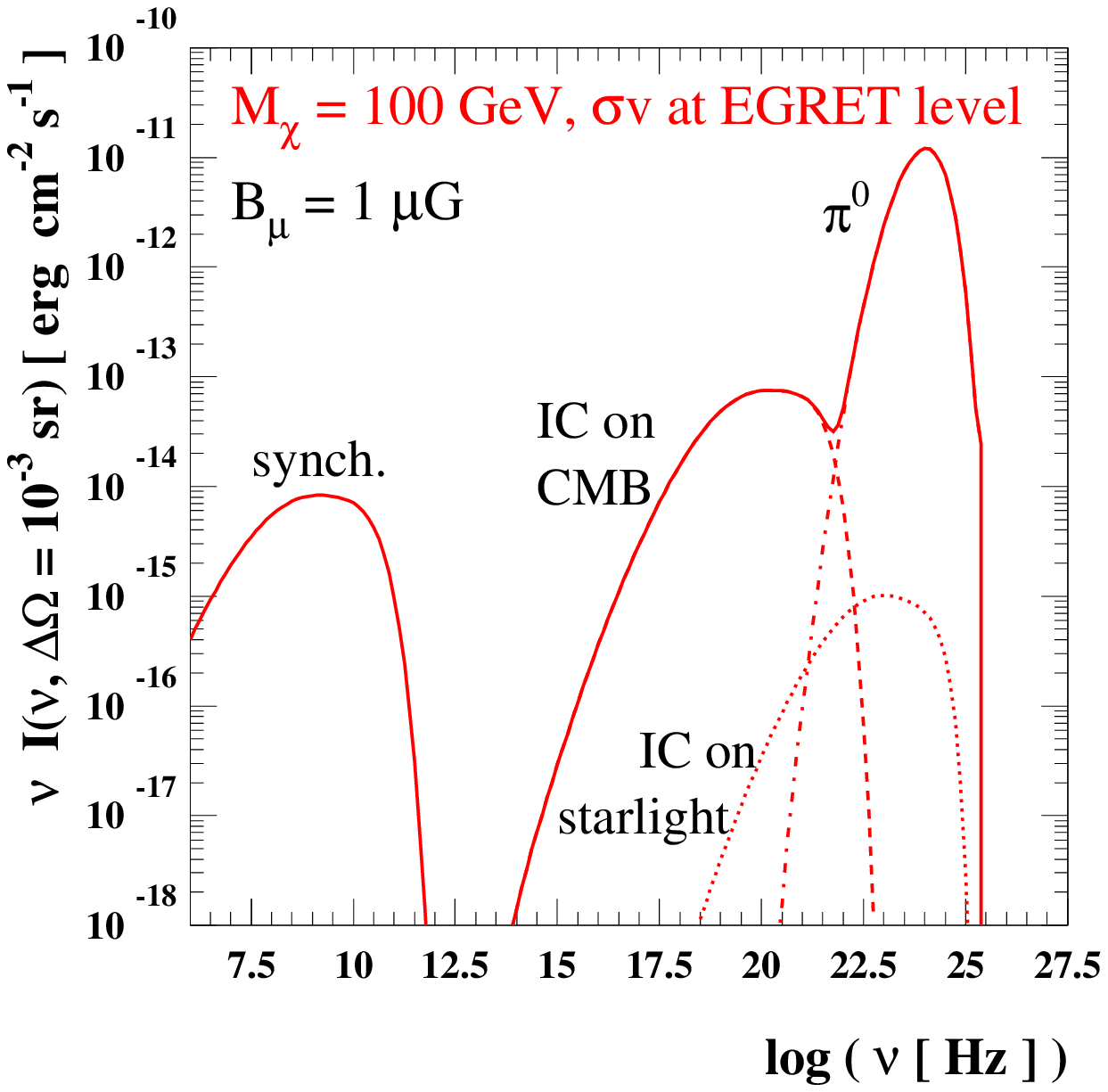}
\quad\includegraphics[scale=0.55]{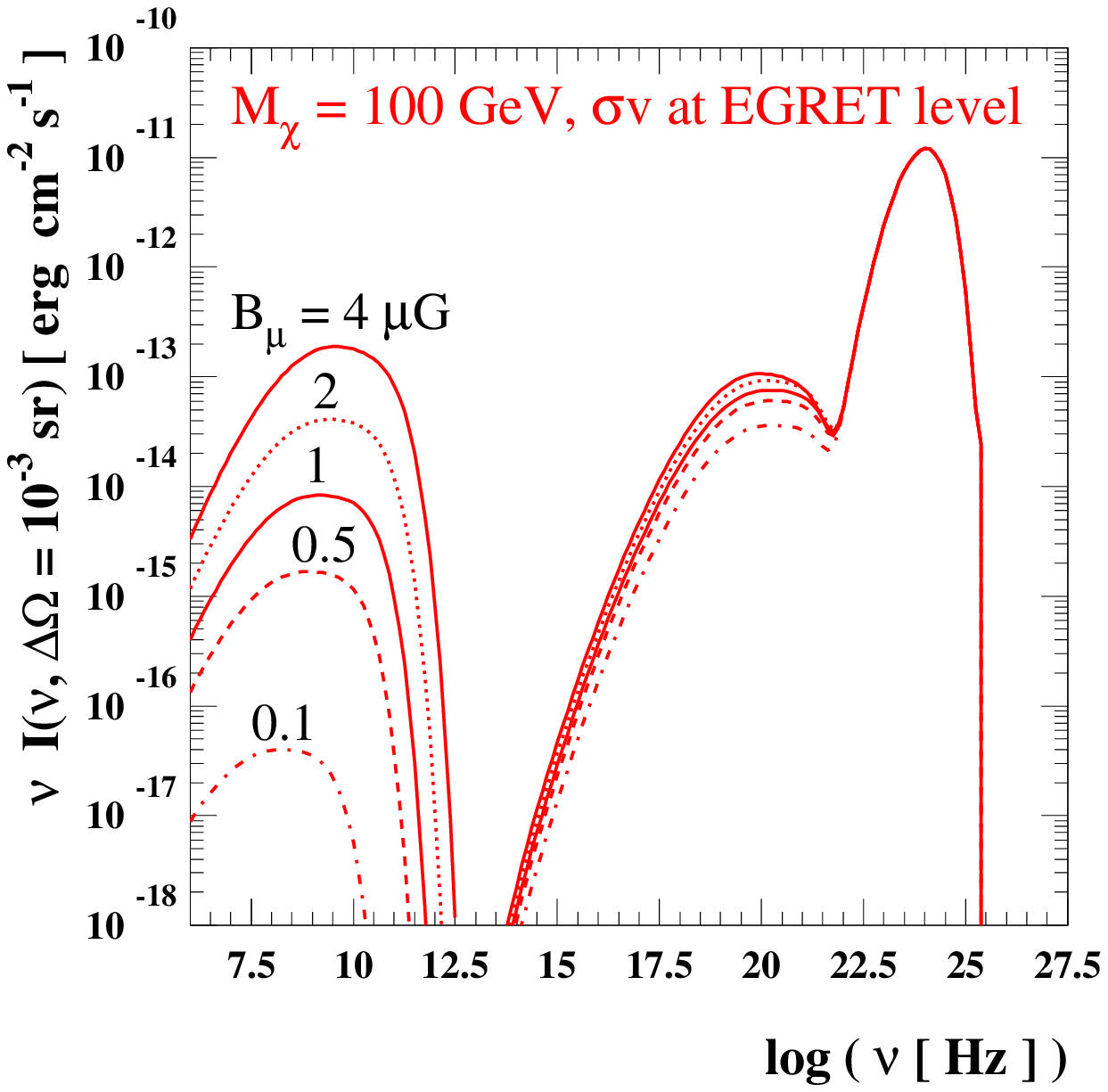}\\
\end{center}
\caption{Detailed multi-wavelength spectrum for a 100 GeV WIMP annihilating into
$b-\bar{b}$ (left), and the effect of varying the magnetic field strength. The WIMP pair annihilation rate
has been tuned to give a gamma-ray signal at the level of the EGRET measured
flux.}
\label{fig:multi1}
\end{figure*}

In Fig.~\ref{fig:multi1} we plot the sample multi-frequency seed of the emission in Draco due
to WIMP annihilations, implementing our reference NFW halo model and reference values
for the WIMP mass, the magnetic field and the various propagation parameters. The WIMP pair annihilation rate
has been tuned to give a gamma-ray signal at the level of the EGRET measured
flux; the displayed surface brightness is in the direction of the center of Draco and for an
angular acceptance equal to the EGRET angular resolution, i.e. not optimized for future
observations (we should have in fact considered different solid angles at different wavelengths).
As apparent, there is a significant component in the X-ray band due to inverse Compton
on the microwave background radiation, while the contribution on starlight is essentially
negligible. Scaling of signals with the assumed value of the magnetic field are also displayed in the right panel.

\begin{figure*}[!t]
\begin{center}
\includegraphics[scale=0.55]{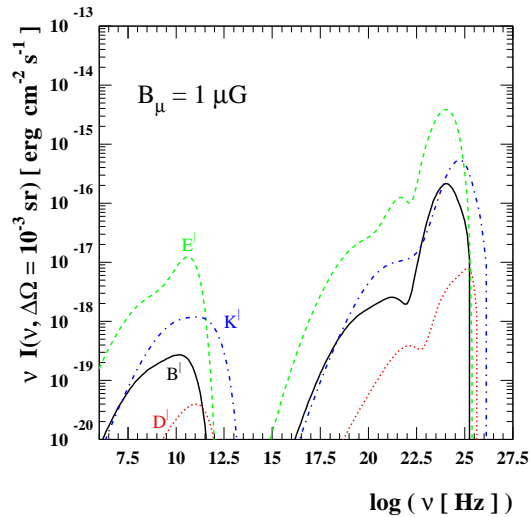} \\
\end{center}
\caption{Multi-wavelength spectra for the four benchmark models described in the text. The prediction
is shown for the best fit NFW profile, and for a mean
magnetic field equal to $1 \mu G$.}
\label{fig:multi2}
\end{figure*}

While so far we have considered simple toy models for the WIMP accounting for the dark
matter halo in Draco, in Fig.~\ref{fig:multi2} we consider a few explicit realizations of
this scenario within the constrained minimal supersymmetric extension to the Standard
Model (cMSSM), picking among the models studied in \cite{Battaglia:2003ab} those better
exemplifying the widest range of possibilities within that particular theoretical setup.
All the models are fully consistent with accelerator and other phenomenological
constraints, and give a neutralino thermal relic abundance exactly matching the central
cosmologically observed value \cite{WMAP}. We adjusted here the values of the universal
soft supersymmetry breaking scalar mass $m_0$ given in \cite{Battaglia:2003ab} in order
to fulfill this latter requirement, making use of the latest Isajet v.7.72 release and of
the DarkSUSY package \cite{ds}. The values of the cMSSM input parameters for the various
models are given in Tab.~\ref{tab:models} (see also Ref.~\cite{Colafrancesco:2005ji}).
Each benchmark model correspond to a different mechanism responsible for the suppression
of the otherwise too large bino relic abundance: ${\bf B}^\prime$ lies in the bulk region
of small supersymmetry breaking masses, and gives a dominant $b-\bar b$ final state;
${\bf D}^\prime$ corresponds to the coannihilation region, and features a large branching
ratio for neutralino pair annihilations in $\tau^+-\tau^-$; ${\bf E}^\prime$ belongs to
the focus point region, with a dominant $W^+-W^-$ final state, and, finally, ${\bf
K}^\prime$ is set to be in the funnel region where neutralinos rapidly annihilate through
$s$-channel heavy Higgses exchanges, dominantly producing $b-\overline b$ pairs as
outcome of annihilations. Not unlike what we found in the case of the multi-wavelength
analysis of neutralino annihilations in the Coma cluster (see fig.~25 in
Ref.~\cite{Colafrancesco:2005ji}), the most promising among the four benchmark models of
Tab.~\ref{tab:models} is model ${\bf E}^\prime$, featuring a large pair annihilation
cross section to begin with; the less promising model is instead model ${\bf D}^\prime$,
for which the mechanism suppressing the neutralino relic abundance in the Early Universe,
stau coannihilations, is not associated to pair annihilations of neutralinos today.

\begin{table}[!b]
\begin{center}
\begin{tabular}{l|c|c|c|c|c|}
{\bf Model}&$M_{1/2}$&$m_0$&$\tan\beta$&${\rm sign}(\mu)$&$m_t$\\ \hline ${\bf B}^\prime$ ({\em
Bulk})& 250 & 57 & 10  & $>0$ & 175\\ ${\bf D}^\prime$ ({\em Coann.})& 525 & 101 & 10  & $>0$ & 175\\
${\bf E}^\prime$ ({\em Focus P.})& 300 & 1653 & 10  & $>0$ & 171\\ ${\bf K}^\prime$ ({\em Funnel})&
1300 & 1070 & 46  & $<0$ & 175\\ \hline
\end{tabular}
\end{center}
\caption{The input parameters of the four cMSSM benchmark models we consider here. The units for the
mass parameters are GeV, and the universal trilinear coupling $A_0$ is set to 0 for all models (see
\cite{Battaglia:2003ab,Colafrancesco:2005ji} for details). }\label{tab:models}
\end{table}

Lastly, we found that the SZ effect produced by DM annihilation in Draco, even though is
a definite probe of the DM annihilation in such cosmic structures (see, e.g.,
\cite{Colafrancesco2004,CEC2006}) is quite low when we take into account the spatial
diffusion of secondary electrons: we find, in fact, that the SZ signal towards the center
of Draco is negligible even when we normalize the gamma-ray signal at the level of the
EGRET upper limit.

\section{A black hole at the center of Draco?}
 \label{setc.BH}

There is one further effect which could change substantially our picture: if a black hole
is present at the center of Draco, and had it formed through an adiabatic accretion
process, the ambient dark matter population would have experienced a sharp increase in
its density profile, turning into a ``spike'' of dark matter, with a dramatic enhancement
in the dark matter annihilation rate at the center of Draco. Such spike was originally
proposed for the Milky
Way~\cite{Gondolo:1999ef,Ullio:2001fb,Merritt:2002vj,Bertone:2005hw} in connection its
central black hole, which has a mass of about $3\, 10^6 \,\msun$, and, more recently, it
has been extrapolated to small mass halos~\cite{Zhao:2005zr,BZS}, including substructures
within the Milky Way dark matter halo, eventually embedding black holes of intermediate
mass, in the range between $10^2$ to $10^6 \msun$.

\begin{figure*}[!t]
\begin{center}
\includegraphics[scale=0.55]{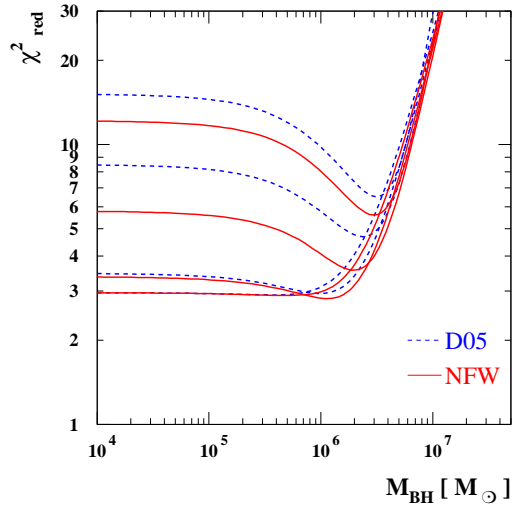}
\end{center}
\caption{Reduced $\chi^2$ for a fit of the star velocity dispersion data of Draco
in the binning of Munoz et al. under the hypothesis that a black hole of given mass is
present at the center of Draco; the dark matter profiles are described by a NFW or a D05
profile with scale factor $a = 1$~kpc and, respectively, $\rho^\prime$ equal to
$10^7$, $2 \cdot 10^7$, $3 \cdot 10^7$, $3.72 \cdot 10^7  \msun$~kpc$^{-3}$
(from top to bottom in the figure;
the last value corresponds to best fit in the case without the black hole) and
$5 \cdot 10^6$, $10^7$, $2 \cdot 10^7$, $2.54 \cdot 10^7  \msun$~kpc$^{-3}$
(again, from top to bottom in the figure, with last value being the best fit in the case
without the black hole).} \label{fig:bh}
\end{figure*}

There is a strong observational evidence for the existence of super-massive black holes
(in mass range between $10^6$ and $10^9 \msun$), without however a detailed understanding
on how those objects form, or on the mechanism enforcing the observed correlations with
properties of the hosting halos. In one of the proposed scenarios, these two issues are
addressed in terms of pre-existing intermediate-mass black hole seeds, forming in turn in
proto-galaxy environments~\cite{islama:2003,volonteri:2003,Koushiappas:2003zn}: a
significant population of these smaller mass objects would still be present in
Galaxy-size halos, most likely associated to substructures which have not been tidally
disrupted, while merging into the halo. Their presence in the Milky Way halo would be
very hard to prove in terms standard astrophysical observations. In particular, there is
no evidence for the presence of a black hole at the center of Draco: it is reasonable to
expect that a black hole, being in such a gas poor environment, would be in a dormant
phase, rather than in an accreting and luminous one. In Fig.~\ref{fig:bh} we sketch the
dynamical response of adding a black hole of given mass on top of the mass models
introduced in Section~\ref{sec:halo} (the response of the dark matter profile, as
specified below, is included): the fit of the star velocity dispersion is not sensitive
to black holes of mass smaller than about $10^5 \msun$, slightly improves for masses
around a few times $10^6 \msun$, while the presence of black hole of mass larger than
about $10^7 \msun$ is dynamically excluded.

We will take a phenomenological approach and make the hypothesis that a black hole of
given mass $M_{BH}$ has formed adiabatically at the center of Draco. The process turns an
initial (i.e. before the black hole has accreted the bulk of its mass) dark-matter
density profile scaling as $\rho_i(r_i) \propto r_i^{-\gamma}$ into a final profile of
the form $\rho_f(r_f) \propto r_f^{-A}$: in a simplified system with all dark matter
particles on circular orbits, it is easy to show that conservation of mass and angular
momentum imply that $A=(9-2\gamma)/(4-\gamma)$ \cite{qhs,Gondolo:1999ef,Ullio:2001fb},
i.e. that the final profile is significantly steeper than the initial; this results holds
also in a general setup. To derive the right normalization, on the other hand, one has to
refer to the full phase space distribution function for the dark matter profile and
implement the appropriate adiabatic invariants. We refer here to the procedure outlined
in~\cite{Ullio:2001fb}; in the same paper it is shown that, since the growth of the spike
depends on the existence of a very large population of cold particles at the center of
the dark matter system, where the black hole is adiabatically growing, large spikes form
for singular profiles, which embed such large number of cold statesr, while it does not
for cored profiles for which  it is not the case. We will discuss then only the case for
the NFW profile and the D05 profile.

\begin{figure*}[!t]
\begin{center}
\includegraphics[scale=0.55]{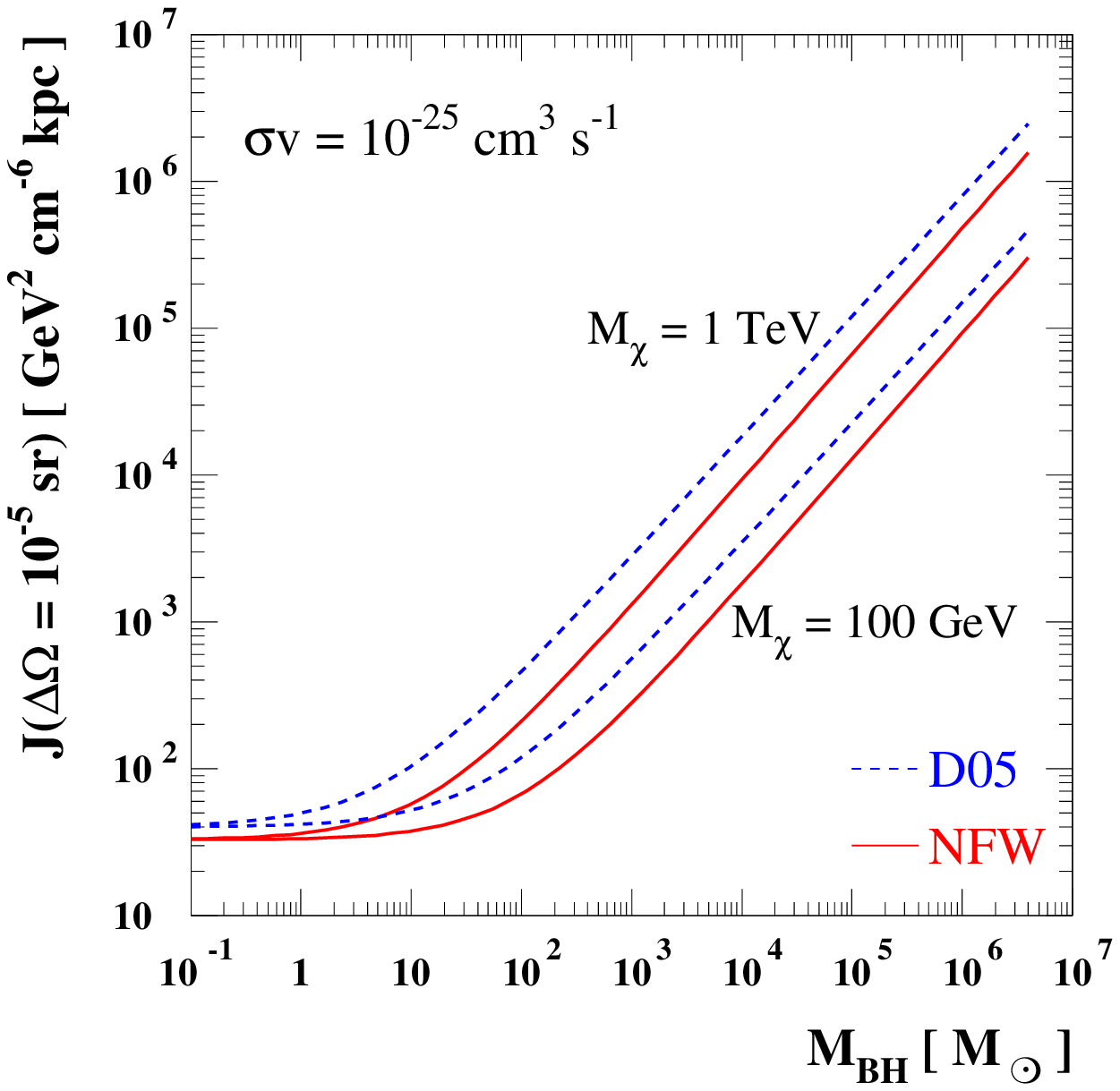}
\quad\includegraphics[scale=0.55]{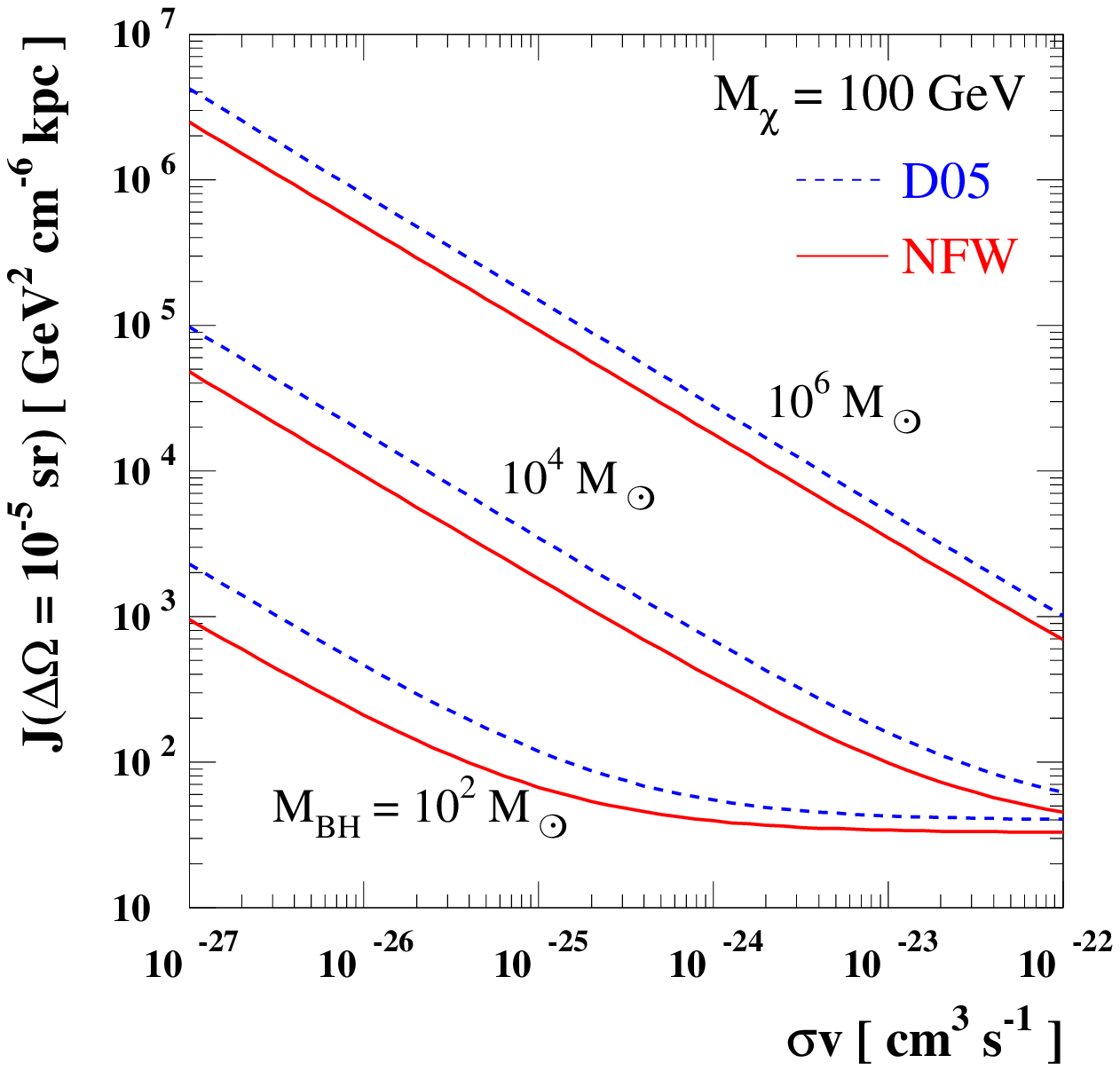}\\
\end{center}
\caption{Left panel:  integral of the  square of dark matter density along the line of
sight towards the center of Draco and averaged over an angular acceptance of
$10^{-5}$~sr, in case of a central  enhancement of the dark matter density due the
adiabatic formation at the center of the system of a black hole of given mass; the
background dark matter profile are those considered in the left panel and for
$\rho^\prime$ equal to $3.72 \cdot 10^7  \msun$~kpc$^{-3}$ (NFW profile) and $2.54 \cdot
10^7  \msun$~kpc$^{-3}$ (D05 profile). The value of the WIMP pair annihilation cross
section and mass enter in the estimate of central maximal WIMP number density set by pair
annihilations. Right panel: same as in the right panel but for a few values of the black
hole mass and varying the pair annihilation cross section.} \label{fig:bh2}
\end{figure*}

In Fig.~\ref{fig:bh2} we plot the line-of-sight integral function $J$, we have introduced
in
Eq.~\ref{eq:los} 
as relevant quantity for predictions on the gamma-ray flux, as a function of the black
hole mass and for a given value of the WIMP annihilation cross section $\sigma v$, or
vice versa. The value of $\sigma v$ enters critically since the very singular spike
density profile has to be extrapolated down to the radius at which a maximal WIMP density
is enforced by WIMP pair annihilations, i.e.~ref{Gondolo:1999ef}:
\begin{equation}
\rho_{\rm max} = \frac{M_\chi}{(\sigma v) (t_0-t_f)}\;,
\end{equation}
where $t_0$ is the present time and $t_f$ the formation time of Draco. As it can be see,
enhancements in the gamma-ray flux of even four orders of magnitude are at hand; scalings
in black hole mass and $\sigma v$ are analogous for the two halo models considered here.

\begin{figure*}[!t]
\begin{center}
\includegraphics[scale=0.55]{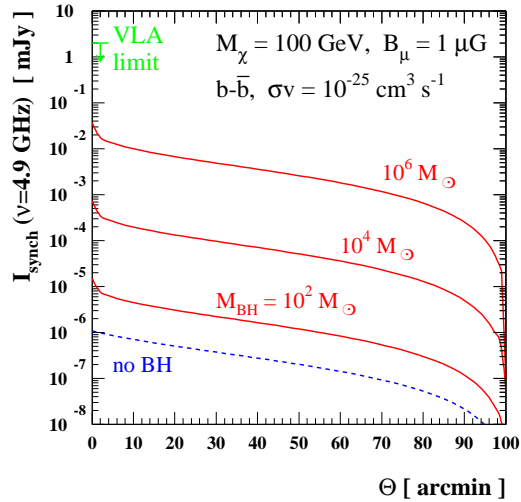}
\end{center}
\caption{We show the radio (at $\nu = 4.9$ GHz) surface brightness of Draco in the case
of an adiabatic growth of a black hole of given mass $M_{BH}= 10^2, 10^4$ and $10^6
M_{\odot}$, as labelled. A reference neutralino $b {\bar b}$ model with $M_{\chi}=100$
GeV and $\sigma v = 10^{-25}$ cm$^3$ s$^{-1}$ with a magnetic field $B_{\mu}=1$ $\mu$G
are adopted here.
 }
 \label{fig:radiolimbh}
\end{figure*}

The spike is confined in a very tiny portion of the halo, essentially the region within
which the black hole dominates the potential well of the final configuration, i.e.
smaller than 1~pc even for the heaviest black holes we are considering. The induced
gamma-ray source would appear as a point source even at future telescopes with improved
angular resolution.  On the other hand, analogously to the effect we have already
discussed for the standard dark matter halo component, the emitted electrons and
positrons diffuse out of the central region and give rise to radio and Inverse Compton
signals on a very wide angular size. In Fig.~\ref{fig:radiolimbh} we consider, for a few
sample masses for the central black hole and one reference annihilation cross section,
the induced radio surface brightness in the same configuration displayed in
Fig.~\ref{fig:radio} (propagation parameters in set \#1).

\begin{figure*}[!t]
\begin{center}
\includegraphics[scale=0.55]{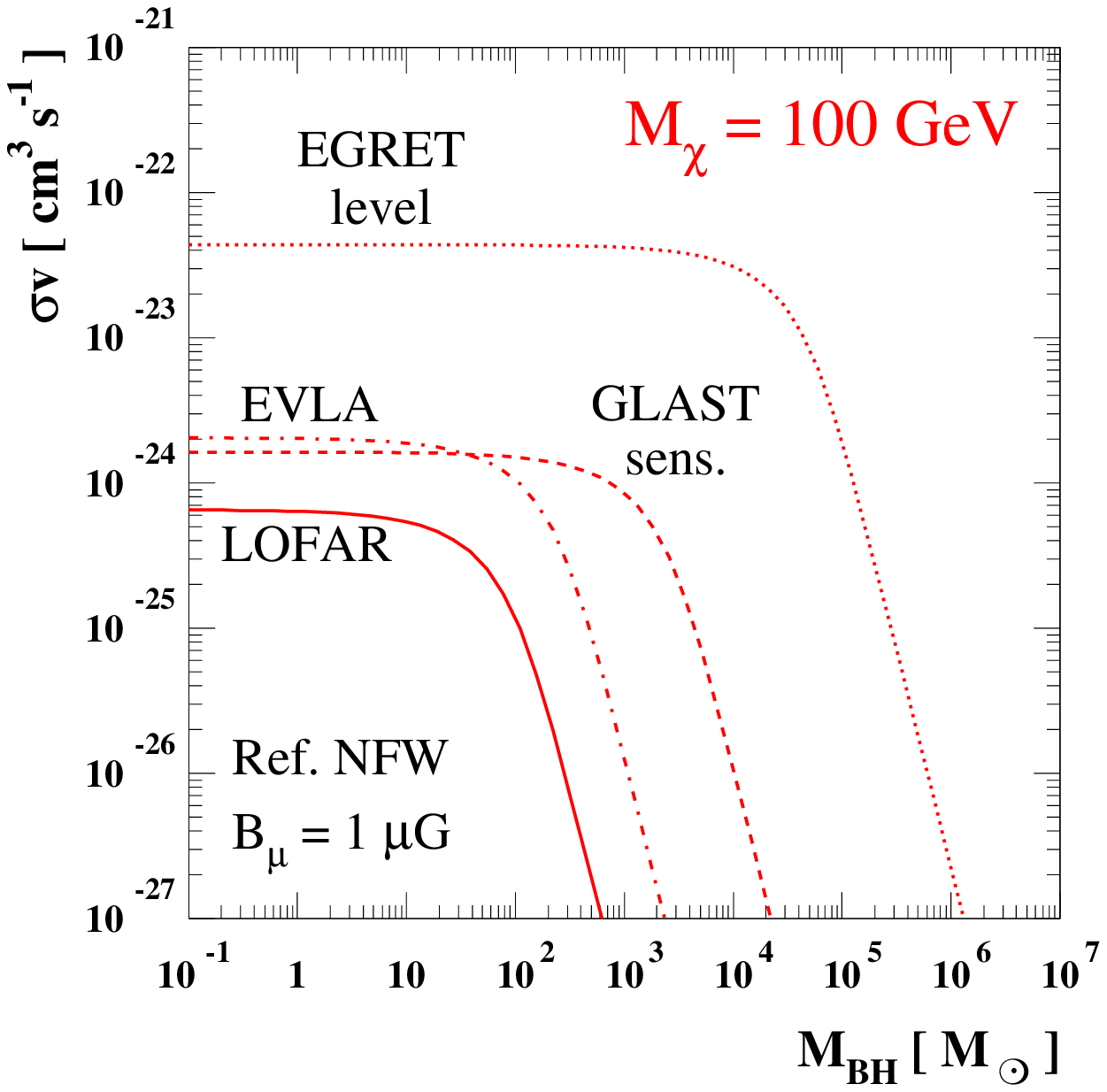}
\quad\includegraphics[scale=0.55]{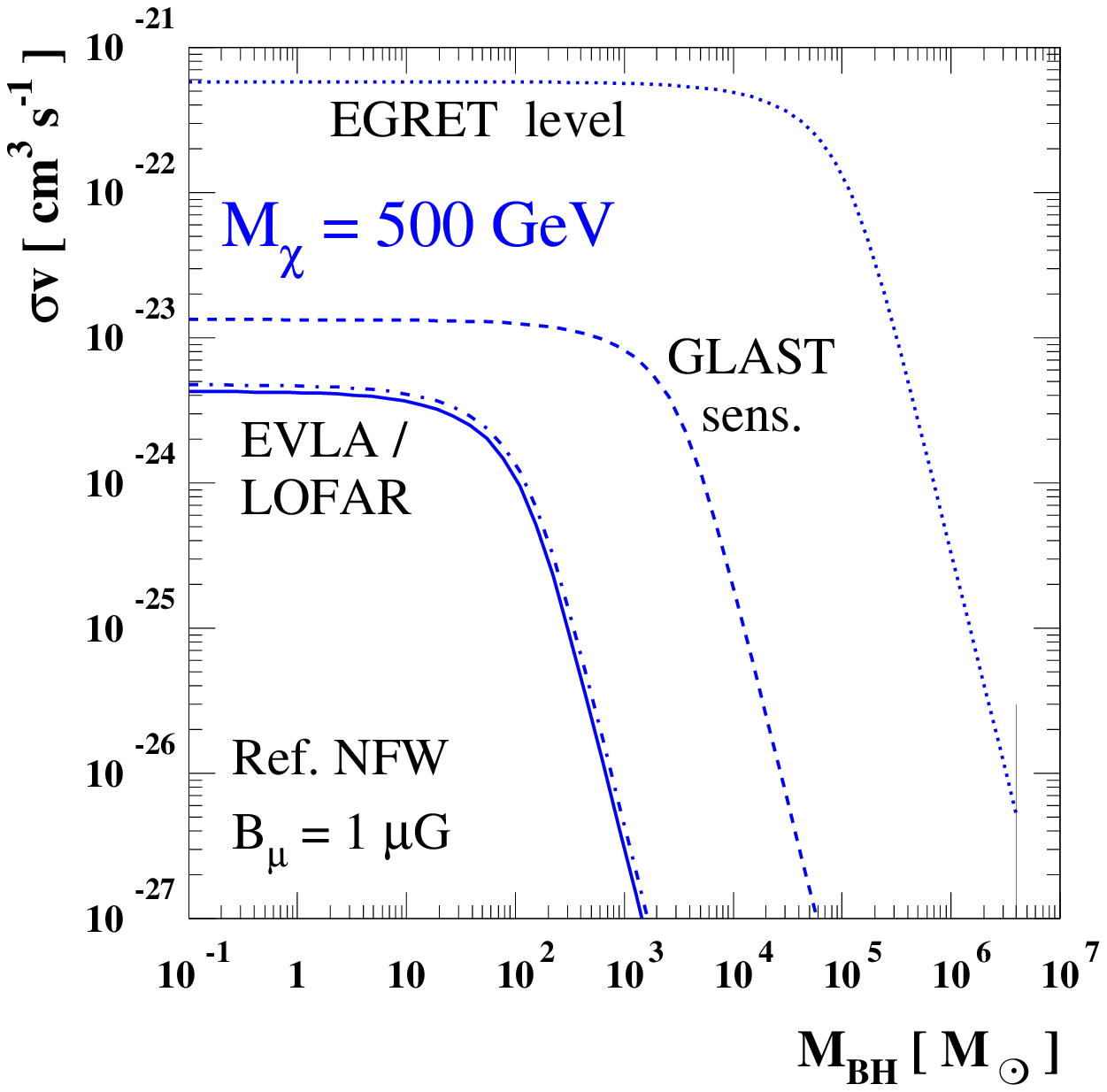}\\
\end{center}
\caption{We show the projected sensitivity of radio (LOFAR, EVLA) and gamma-ray (EGRET,
GLAST) experiments observing Draco as a function of the black hole mass $M_{BH}$ for two
choices of the neutralino mass: $M_{\chi}=100$ GeV (left panel) and $M_{\chi}=500$ GeV
(right panel). A reference case of a NFW dark matter profile and a magnetic field
$B_{\mu}=1$ $\mu$G are adopted here.} \label{fig:radiolimbh2}
\end{figure*}

In Fig.~\ref{fig:radiolimbh2}  we show the scaling of future expected sensitivities in
the plane annihilation rate versus black hole mass,  for two sample WIMP masses,  the
$b-\bar{b}$ annihilation channel, the reference NFW halo profile and set of propagation
parameters. As already mentioned above, the effect of the adiabatic black hole growth is
more dramatic for WIMP models with smaller annihilation cross section; for comparable
annihilation cross section the enhancement in the signal is larger at radio wavelengths
than for gamma-rays. There is a 1/4$\pi$ mismatch since we are essentially adding a point
source at the center of the system: we detect only the gamma-rays emitted in our
direction, while all emitted electrons and positrons pile up into the population giving
rise to the radiation at lower frequencies.

\section{Conclusions}

Following a program of thorough investigation of the multi-wavelength yields of WIMP dark
matter annihilations started with the case of the Coma cluster in Ref.~\cite{Colafrancesco:2005ji}, in this paper we analyzed the case of the nearby dSph
Draco. Under the assumption of equilibrium for the stellar component, we made use of the
large wealth of available photometric studies to derive precise mass models for Draco.
Under a general setup, we studied the preferred values for the dark matter halo length
scale, its density normalization parameter as well as its anisotropy parameter. Results
from numerical simulations of structure formation, together with a proper treatment of
the effects of tides on the density profile, enabled us to correlate the concentration
parameter and the initial virial mass of the dSph under consideration. In turn, this
allowed us to further constrain the best-fit dark matter halo models for Draco.

We then proceeded to an evaluation of the gamma-ray and electron/positron yield expected
from Draco under the hypothesis that the dark matter is in the form of a
pair-annihilating WIMP. To this extent, we resorted to illustrative cases of WIMPs of
given mass and pair-annihilation cross section, dominantly annihilating into final states
giving rise to the two extrema of a soft and a hard photon spectrum. For definiteness,
and for illustrative purposes, we also considered theoretically well motivated benchmark
supersymmetric models.

We pointed out that unlike the case of the Milky Way galactic center, the spread in the
estimate of the gamma-ray flux from Draco is significantly narrow, once the particle
physics setup for the dark matter constituent is specified, and that Draco would appear
as a point-like gamma-ray source in both ACTs and GLAST observations. In analogy with our
procedure carried out in CPU2006~\cite{Colafrancesco:2005ji} for larger dark matter
halos, we implemented a fully self-consistent propagation setup for positrons and
electrons produced in WIMP pair annihilations, and we studied the subsequent generation
of radiation in the radio frequencies from synchrotron emissions, and at higher
frequencies from inverse Compton scattering off starlight and cosmic microwave background
photons.

We showed that, unlike in larger dark matter halos, as it is the case for the Coma
cluster~\cite{Colafrancesco:2005ji}, in small, nearby objects the spatial diffusion of
electrons and positrons plays a very significant role. As a consequence, the expected
radio emission from Draco is spatially extended, and, depending upon the propagation
setup and the values of the magnetic field in Draco, can provide a detectable signal for
future radio telescopes. In some cases, we find that an extended radio emission could be
detectable from Draco even if no gamma-ray source is identified by GLAST or by ACTs,
making this technique the most promising search for dark matter signatures from the class of
objects under consideration, i.e. nearby dwarf spheroidal galaxies.

We finally showed that available data can accommodate the presence of a black hole in the
center of Draco, even improving the fit to the data for some values of the black hole
mass. The corresponding expected enhancement in the gamma-ray flux and in the radio
surface brightness for cuspy dark matter halo profiles and an adiabatic growth of the
black hole can be of several orders of magnitude. If the mass of the black hole is around
or larger than $10^6 M_{\odot}$, WIMP models are predicted to give unmistakable
astrophysical signatures both for future gamma-ray telescopes and for future radio
telescopes.



\begin{thebibliography}{99}
\bibitem{glastsens}
See {\tt www-glast.slac.stanford.edu/software/IS/glast\_lat\_performance.htm}.
\end{thebibliography}
\end{document}